\documentclass[12pt, a4paper]{article}

\usepackage[utf8]{inputenc} 
\usepackage{authblk} 

\title{\textbf{Analysis of lattice locations of deuterium in tungsten and its application for predicting deuterium trapping conditions}}

\author[1]{Xin Jin}
\author[1]{Flyura Djurabekova}
\author[2]{Etienne A. Hodille}
\author[3]{Sabina Markelj}
\author[1]{Kai Nordlund}

\affil[1]{\normalsize\textit{Department of Physics, University of Helsinki, P.O. Box 43, Helsinki FIN-00014, Finland}}
\affil[2]{\normalsize\textit{CEA Cadarache, IRFM, F13108 Saint Paul Lez Durance, France}}
\affil[3]{\normalsize\textit{Jožef Stefan Institute (JSI), Ljubljana, Slovenia}}

\date{}
\usepackage[pdfborder={0 0 0}]{hyperref} 
\usepackage{graphicx} 
\usepackage{float} 
\usepackage{subcaption} 
\usepackage[labelfont=bf]{caption} 
\usepackage{indentfirst} 
\usepackage[top=1.2in, bottom=1.25in, left=0.9in, right=0.9in]{geometry}
\usepackage{amsmath} 
\usepackage{amssymb}
\usepackage{bm} 
\usepackage[T1]{fontenc} 
\usepackage{mathtools}
\usepackage{siunitx} 
\usepackage{tabularx} 
\usepackage{booktabs} 
\usepackage[sort&compress, numbers]{natbib} 
\usepackage{xcolor}

\DeclareUnicodeCharacter{2010}{-}
\DeclareUnicodeCharacter{01E3F}{\'{e}}

\begin{document}
\maketitle

\section*{Abstract}

Retention of hydrogen isotopes (protium, deuterium and tritium) in tungsten is one of the most severe issues in design of fusion power plants, since significant trapping of tritium may cause exceeding radioactivity safety limits in future reactors. Hydrogen isotopes in tungsten can be detected using the nuclear reaction analysis method in channeling mode (NRA/C). However, the information hidden within the experimental spectra is subject to interpretation. In this work, we propose the methodology to interpret the response of the experimental NRA/C spectra to the specific lattice locations of deuterium by simulations of the NRA/C spectra from atomic structures of deuterium lattice locations as obtained from the 
first principles calculations. We show that trapping conditions, i.e., states of local crystal structures retaining deuterium, affect the lattice locations of deuterium and the change of lattice locations can be detected by ion channeling method. By analyzing the experimental data, we are able to determine specific information on the deuterium trapping conditions, including the number of deuterium atoms trapped by one vacancy as well as the presence of impurity atoms along with deuterium in vacancies.  

\section{Introduction}

The interaction of hydrogen isotopes (protium, deuterium and tritium) with metals has garnered sustained interest due to application of hydrogen in energy storage \cite{Sakintuna2007} as well as its adverse effects on the properties of structural materials \cite{Song2013, Tiegel2016}. Among various metals, the interaction of hydrogen isotopes with tungsten has received particular attention, in particular because tungsten is a promising candidate for the plasma facing wall of fusion reactors \cite{Tanabe2014}, where it needs to withstand extremely high fluxes of hydrogen isotopes \cite{Roth2011, Rieth2019}. A comprehensive understanding of the behavior of hydrogen isotopes in tungsten, including its lattice locations and trapping conditions, is essential for the development of future fusion devices.
  
  \bigbreak
  
Since hydrogen is the lightest element, identifying its geometrical configuration in the crystal structure of materials is not a straightforward process. For example, direct imaging of hydrogen in metal hydride is a demanding task due to the weight difference between hydrogen and heavy elements \cite{Graaf2020}. The low mass of hydrogen renders it essentially invisible to ion beam analysis methods that rely on backscattering techniques. On the other hand, the low atomic number of hydrogen opens the gateway for methods that take advantage of its weak Coulomb barrier, i.e., nuclear reaction analysis (NRA) \cite{Bubert2002}. NRA has been used to study hydrogen in metals over decades \cite{Picraux1974a, Amsel1984, Myers1989, Tor94,Alimov2001, Markelj2014}. By using a D($^3$He,p)$^4$He reaction, it can perform a depth profiling  of deuterium in tungsten up to several micrometers, with a depth resolution on the order of magnitude of submicrometer \cite{Mayer2009}. 

\bigbreak

 Notably, NRA in channeling mode (NRA/C) is a unique method to determine the lattice locations of deuterium in tungsten \cite{Feldman1982}. In this mode, probing ions are confined in open channels between rows of target atoms so that NRA/C signals would depend on the location of impurity atoms to be studied. By applying NRA/C, Picraux and Vook \cite{Picraux1974, Picraux1975} investigated the lattice locations of 15 - 30 keV deuterium implanted into tungsten at room temperature with a relatively low fluence ($\sim 10^{15}$ cm$^{-2}$). They performed angular scans through <100> axis of tungsten, and concluded that the deuterium atoms were located at a position at or close to tetrahedral interstitial sites (TIS). Ligeon \textit{el al.} \cite{Ligeon1986} explored the location of deuterium in a large group of metals. For the case of tungsten, they implanted 10 - 15 keV deuterium and suggested that the occupation site of deuterium was slightly displaced from the TIS in a temperature range of 15 K to room temperature. Nagata \textit{et al.} \cite{Nagata2000} carried out similar NRA/C experiments using 1 keV deuterium, and reported similar results.   
  
  \bigbreak
  
In principle, the lattice locations of deuterium should be affected by the trapping conditions. This trapping condition defines by which kind of crystal structures the hydrogen atoms are trapped, for example, monovacancy or vacancy clusters. It also describes the actual state of the trapping structures, for example, how many hydrogen atoms are in a monovacancy and if there are impurity atoms in the monovacancy. The trapping conditions of hydrogen in tungsten have been extensively studied by first principles calculations based on density functional theory (DFT). Note that the trapping conditions of different hydrogen isotopes should be similar, as the isotope mass does not affect chemical bonding, and hence DFT calculations or experiments for protium are directly applicable to deuterium, and vice versa. Several independently performed DFT calculations have shown that hydrogen prefers to occupy the tetrahedral interstitial site in a perfect tungsten \cite{Henriksson2005, Liu2009, Zhou2012}. However, the solubility of hydrogen in tungsten is very low \cite{Lu2014} so that the hydrogen retained in tungsten at room temperature is mainly trapped by crystal defects \cite{Tanabe2014}. Moreover, experiments of hydrogen and deuterium diffusion have shown that single hydrogen or deuterium atoms migrate very rapidly in tungsten even at cryogenic temperatures via a quantum-mechanical migration mechanism \cite{Tanabe2014,wipf2014hydrogen,Frauenfelder1969,Fly70}. Hence it is implausible that any appreciable concentration of isolated hydrogen or deuterium atoms would be present in pristine tungsten in experiments. Instead, The trapping of hydrogen by vacancy-type defects, such as, vacancy and vacancy clusters, is of particular interest due to high binding energies (> 1 eV) \cite{Johnson2010, Kato2011}. Furthermore, the open volume of vacancy-type defects is able to accommodate multiple hydrogen atoms \cite{Hou2019}. Interstitial-type defects can also attract hydrogen, but the corresponding binding energies tend to be weak \cite{Becquart2010, Backer2017}.
  
  \bigbreak
  
The insights gained from DFT calculations have frequently been used to interpret trapping conditions of deuterium in experiments \cite{Guterl2015, Ogorodnikova2015, Zibrov2016}. However, with regard to the lattice locations of deuterium, a connection between DFT calculations and NRA/C experiments has not been well established. In fact, results obtained from the two methods are even seemingly conflicting with each other. As an example, if we assume that primary trapping structures for the deuterium in the previously mentioned NRA/C experiments \cite{Picraux1974} are vacancies as postulated in literature \cite{Picraux1981}, then we have a contradictory result: the NRA/C experiments indicate that deuterium atoms are located near the TIS; whereas DFT calculations indicate that the lattice locations of deuterium is near the octahedral interstitial site (OIS) \cite{Liu2009a, Heinola2010a, Johnson2010, Sun2013}.
  
\bigbreak  
  
This apparent paradox indicates that deuterium trapping conditions possess a complex nature. The change of exact state of trapping conditions can affect the lattice locations of deuterium. This follows from the DFT calculations which show that the hydrogen atoms tend to shift from OIS to TIS position in a vacancy, when the filling level of hydrogen, i.e., the number of hydrogen atoms trapped by the vacancy, increases up to 12 \cite{Ohsawa2010}. Yet, the temperature effect imposes an additional complication: it is reported that at room temperature a vacancy can hold in only up to 6 hydrogen atoms \cite{Sun2013, Fernandez2015}. Hence, it is not clear whether the decoration of vacancies by no more than 6 deuterium atoms can be used to explain the NRA/C experimental results.

\bigbreak
  
In this work, we investigated the lattice locations of deuterium in tungsten by combination of NRA/C methods and DFT calculations. DFT calculations were employed to determine the lattice locations of hydrogen (including its isotope deuterium) in various trapping conditions. In order to establish a robust connection between DFT calculations and NRA/C experiments, we developed an NRA/C simulation tool which uses the DFT results as an input and generates the NRA/C signals comparable to experimental ones. We present the response of the simulated NRA/C signals to deuterium trapped at different conditions and compare the simulated spectra to the experimental ones from Refs.\cite{Picraux1974, Picraux1975}. We demonstrate that the connection between the lattice locations and trapping conditions of deuterium can not only be used to resolve the inconsistency between the NRA/C experiments and DFT calculations, but also can provide valuable information regarding the filling level of deuterium in vacancies and evolution of trapping conditions. 
  
\section{Methodology}

\subsection{First principles calculations} \label{sec:DFT_method}
  
First principles calculations based on the DFT approach were performed using the Vienna \emph{ab initio} simulation package (VASP) \cite{Kresse1993, Kresse1994, Kresse1996} with the projector augmented wave potentials \cite{Blochl1994, Kresse1999}. The electron exchange correlation was described with generalized gradient approximation (GGA) using Perdew-Burke-Ernzerhof (PBE) functionals \cite{Perdew1996}. The configuration of valence electrons used for tungsten, hydrogen, helium, carbon and oxygen were $5d^5 6s^1$, $1s^1$, $1s^2$, $2s^2 2p^2$ and $2s^2 2p^4$, respectively. Since these DFT calculations seek the ground state configuration of atoms at 0 K, the results will be the same for any hydrogen isotope.
We employed a $3 \times 3 \times 3$ supercell composed of 54 lattice points, and applied a $5 \times 5 \times 5$ $k$-point grid obtained using the Monkhorst-Pack method \cite{Monkhorst1976}. The lattice parameter of the perfect tungsten determined by these DFT calculations is 3.172 Å.  A plane wave energy cutoff of 350 eV was used for most of the calculations. When the supercell contained helium or oxygen, the energy cutoff was increased to 500 eV. Relaxation of atomic positions was performed, in which the energy and atomic force convergence criteria were set to $1 \times 10^{-5}$ eV and $2 \times 10 ^{-3}$ eV/Å, respectively. After the relaxation, the positions of hydrogen atoms were transferred to the NRA/C simulations. 
  
\bigbreak
  
The binding energy between two entities $A_1$ and $A_2$, $E_b^{A_1,A_2}$, is calculated as follows:
  \begin{equation}
  E_b^{A_1,A_2} = (E^{A_1} + E^{A_2}) - (E^{A_1 + A_2} + E_{\text{ref}}) \label{Eq:bind_ener1}
  \end{equation}  
where $E^{A_1}$ and $E^{A_2}$ represent the energy of supercell containing $A_1$ and $A_2$, respectively, $E^{A_1 + A_2}$ represents the energy of supercell containing both $A_1$ and $A_2$ in interaction with each other, and $E_{ref}$ represents the energy of the supercell without $A_1$ and $A_2$. Since zero point energy (ZPE) is not negligible for light atoms, ZPE calculations were performed for hydrogen and helium atoms based on harmonic approximations, in which the positions of other atoms were kept fixed. The binding energy including the ZPE correction, $E_{b,\text{ZPE}}^{A_1,A_2}$, is calculated by:
   \begin{equation}
  E_{b,\text{ZPE}}^{A_1,A_2} = E_b^{A_1,A_2} + [\text{ZPE}(A_1) + \text{ZPE}(A_2)] - \text{ZPE}(A_1 + A_2)\label{Eq:zpe1}
  \end{equation}  
where $\text{ZPE}(A_1)$, $\text{ZPE}(A_2)$ and $\text{ZPE}(A_1+A_2)$ are the total ZPE of suppercell containing $A_1$, $A_2$ and the $A_1 A_2$ complex, respectively. 

\bigbreak

The formation energy of the entity $A_1$, $E_f^{A_1}$, is calculated as follows:
  \begin{equation}
  E_f^{A_1} = E^{A_1} - (E_{\text{ref}} + \sum_i n_i E_i)\label{Eq:form_ener1}
  \end{equation}
where $n_i$ is the difference in the number of element $i$ between the supercell with and without $A_1$, and $E_i$ represents the energy of element $i$ (for hydrogen, $E_i$ is taken as the half of the energy of a hydrogen molecule). Using a similar method to Eq.(\ref{Eq:zpe1}), the formation energy including ZPE correction, $E_{f,\text{ZPE}}^{A_1}$, is calculated by adding the corresponding ZPE to each term of Eq.(\ref{Eq:form_ener1}).

  \subsection{Development of NRA/C simulation program}
    
  We developed a NRA/C simulation program based on a Monte Carlo code called RBSADEC that was originally developed to simulate Rutherford backscattering spectrometry in channeling mode (RBS/c) \cite{Zhang2016, Jin2020, RBSADEC_ND}. An incorporation of NRA/C and RBS/c in the same code is natural due to the intimate relation between the two techniques \cite{Nastasi1995}. The simulation deals with nuclear reactions in the following form:
  \begin{equation}
  a + X \longrightarrow b + Y \pm Q \label{Eq:nucReaction}
  \end{equation}
  where $a$ is a probing ion, $X$ is a target atom that can have nuclear reaction with $a$, $b$ is an emitting particle that is to be detected, $Y$ is a residual nucleus, and $Q$ represents an energy balance. 
  
  \bigbreak
  
  As in the RBS/c counterpart, NRA/C signals are generated as probing ions penetrating through a target that can contain arbitrary atomic structures \cite{Zhang2016}. The trajectory of probing ions is determined by the interaction of probing ions with target atoms based on the binary collision approximation \cite{Robinson1974}. The interactions are calculated using the Ziegler–Biersack–Littmark (ZBL) universal interatomic potential \cite{Ziegler2015}.
  
  \bigbreak
  
  Due to low cross sections, the probability of having nuclear reactions is very low, for which the probing ion must have an extremely close encounter with target atoms. For the particular nuclear reaction considered in this work, the differential cross section is of the order of magnitude of 0.01 barn per steradian \cite{Wielunska2016}, which means that the impact parameter, $b_r$, should be on the order of magnitude of 1 fm. To accelerate the simulation process, the occurrence of a nuclear reaction is determined by the so-called nuclear encounter probability, $P_E$, as follows \cite{Smulders1994, Zhang2016} :
  \begin{equation}
  P_E = \frac{1}{ 2 \pi u_1^2} e^{- b_r^2 / (2 u_1^2)} \label{Eq:encounter_prob}
  \end{equation}
  where $u_1$ represents the one-dimension (1D) thermal vibration magnitude of the target atoms.

  \bigbreak
  
  After each nuclear reaction, the emitting particle is set to move towards a detector outside the target. The energy of the emitting particle is calculated based on the conservation of total energy and linear momentum \cite{Mayer1977}. The calculation method is applied to probing ions of the order to 10 MeV or less, above which mesons or other exotic particles can be produced \cite{Krane1987}.
  
  \bigbreak
  
  During the passage of the emitting particle to the detector, the target is considered to be amorphous in order to further accelerate the simulation. A main difference with RBS/c simulations is that the types of the probing ion and emitting particle are usually different in NRA/C. Thus, different stopping powers are applied to the probing ion and emitting particle. Otherwise, the movement of the emitting particle is taken into account in a similar way to that of RBS/c counterpart, a detailed description of which can be found in previous works \cite{Zhang2016, Jin2020}. Once the emitting particle reaches the detector, its contribution to NRA/C signals is weighted by a differential nuclear reaction cross section. In addition, multiple scattering can be a major source affecting the depth resolution of NRA \cite{Mayer2009}. This effect is taken into account by setting the spread angle of emitting particles.

  \bigbreak

  Comparisons of NRA simulations and experiments under non-channeling conditions can be found in the supplementary materials Section 1, which indicates the validity of the program.
  
  \subsection{Ion channeling simulations}
  
  Results of ion channeling simulations, including NRA/C and RBS/c simulations, are mainly compared to the experiments conducted by Picraux and Vook (P-V experiments) \cite{Picraux1974}. Hence, the setup of ion channeling simulations were based on the P-V experiments.

    \subsubsection{Model targets for NRA/C simulations}
    \label{Sec:model_targ}
    
In NRA/C simulations, tungsten targets containing deuterium were constructed according to the irradiation condition in the P-V experiments, in which 30 keV deuterium was implanted to tungsten to a damage dose of $3 \times 10^{15}$ cm$^{-2}$. Since nuclear reactions strongly depend on isotope (contrary to the DFT calculations), all NRA/C modelling was done specifically for the isotope used in the experiment, deuterium. The incident direction was 7$^{\circ}$ off from the <100> direction. In order to obtain information related to the distribution of deuterium in the experimental targets, SRIM \cite{Ziegler_ND} calculations using the Kinchin-Pease mode were performed by setting the displacement threshold energy of a tungsten atom, $E_d$, to 90 eV \cite{Banisalman2017}. 

\bigbreak

We note that under some irradiation conditions, the depth distribution of implanted ions and damage in polycrystalline or single crystalline materials can differ from the SRIM predictions remarkably due to channeling effects \cite{Nor16}. For the current irradiation condition, we evaluated the effect of channeling on value of the mean ion range using the MDRANGE code \cite{Nor94b,Nor16,MDRANGE} (for details see the supplementary materials Section 2). Our estimations resulted in only at most $\sim$ 10\% difference as compared to the SRIM calculations, when the sample is tilted with respect to the ion beam to avoid the channeling directions. Moreover, even if channeling may somewhat enhance the ranges, this would not affect the analysis of lattice locations of D in W. Hence in the remainder of the paper we use the SRIM results only.

\begin{figure}[h!]
  \centering
  \includegraphics[width=0.6\linewidth]{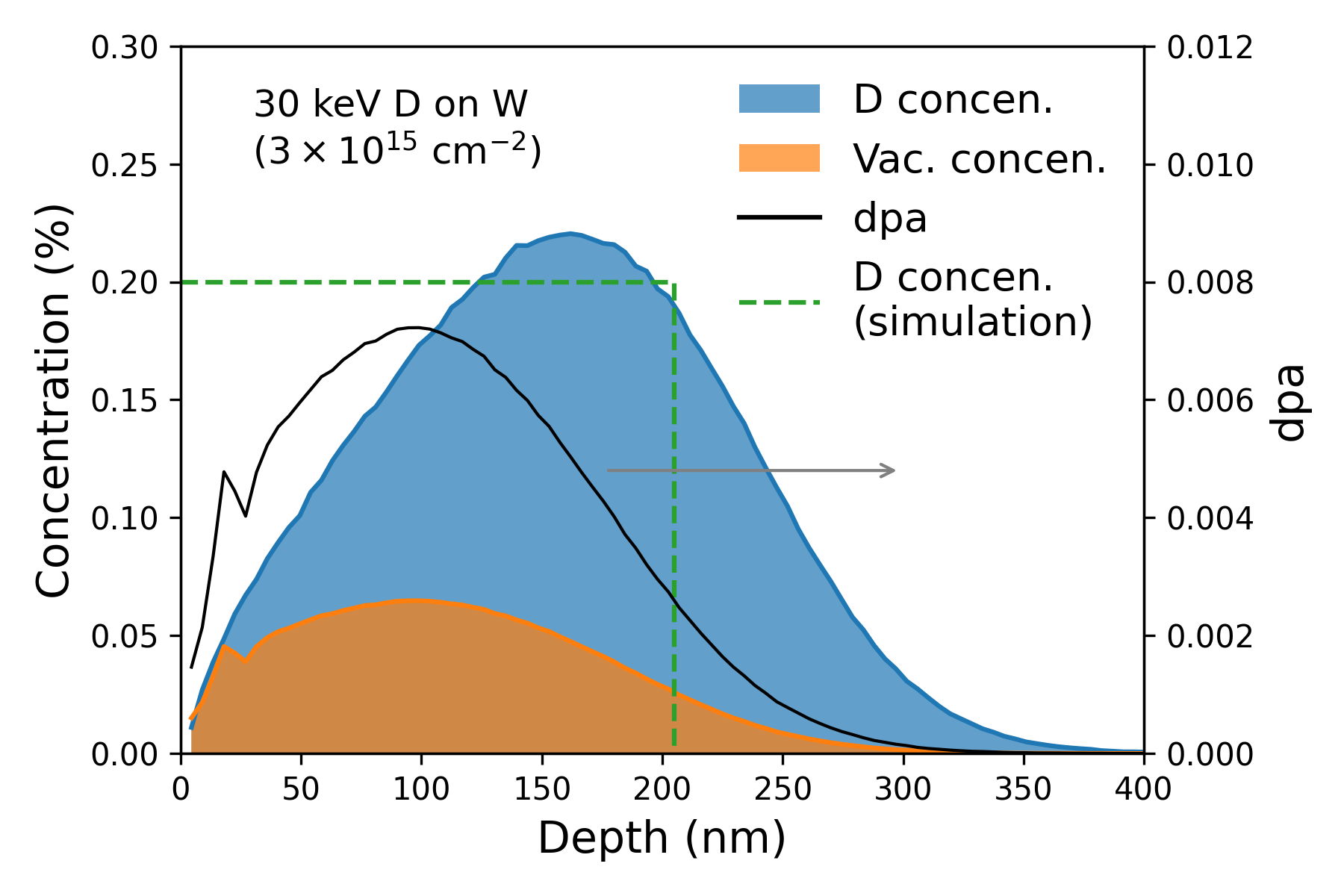}
  \caption{Depth distributions of deuterium concentration (blue) and damage profile in dpa (black) in tungsten irradiated with 30 keV deuterium as calculated by SRIM, depth distributions of vacancy concentration (orange) obtained from the dpa profile and deuterium concentration in simulation targets (green). (The arrow indicates the axis of the damage profile in dpa.) }
  \label{Fig:srim_profile}
\end{figure}

\bigbreak
Fig. \ref{Fig:srim_profile} shows the depth profile of deuterium concentration (blue colored region) and damage (black line) in number of displacements per target atom (dpa) \cite{NRT,Sto13,Nor18} in the experimental targets as calculated by SRIM. The average range of the deuterium is 154.4 nm, and the maximum concentration is 0.22 \%. The maximum damage dose is 0.007 dpa located at 99.0 nm.
  
  \bigbreak
  
  The defect evolution as a function of damage dose (in dpa) in tungsten has been studied by molecular dynamics (MD) simulations performed by Granberg \textit{et al.} \cite{Granberg2021}. According to the results of MD studies, the evolution of vacancy concentration (in \%) with damage dose can be fitted by a function as follows: $0.215 \times [1 - \text{exp}(-49.558 \cdot \text{dpa})]$. Note that the $E_d$ of tungsten in the original MD study was set to 70 eV. Thus, the dpa value in this work is 1.3 times smaller than that in the original MD study. In addition, the number of vacancy obtained from MD simulations is an estimation. Experimental results can differ with that in MD simulation. Using this fitting function, the depth profile of vacancy concentration induced by the 30 keV deuterium irradiation in the experiment was calculated and displayed in Fig. \ref{Fig:srim_profile} (orange colored region). The maximum vacancy concentration is 0.065 \%. The ratio of the total amount of deuterium to the total number of vacancies is 3.8.  
    
  \bigbreak
  
Based on the information obtained from SRIM calculations, the maximum depth of tungsten targets used in the NRA/C simulations was set to 205 nm, which is larger than the average range of 30 keV deuterium (154.4 nm), and is deep enough to cover most of the damaged region. Tungsten atoms were placed at perfect lattice positions. Since the depth resolution of NRA experiments is of the order of 100 nm \cite{Mayer2009}, it is difficult to obtain a depth-resolved NRA experimental signals from a region with a thickness of 205 nm. Thus, the deuterium atoms in the simulation targets were set to have a uniform distribution along the depth as shown in Fig. \ref{Fig:srim_profile} (green line). The concentration of deuterium atoms was set to 0.2 \% (close to the maximum concentration in the experiments).  The lattice locations of deuterium atoms were determined from the DFT calculations. In this process, we take into account that, for every deuterium atom, there are three equivalent positions when it is viewed from the three different <001> directions. Since, as shown in Fig. \ref{Fig:srim_profile}, the maximum vacancy concentration (0.065 \%) induced by the irradiation is fairly small, defective structures were not introduced in the simulation targets. This small amount of defects should not have a significant effect on the movement of probing ions. 
  
  \subsubsection{Setup of NRA/C simulations}
  
  As in the P-V experiments, 750 keV $^3$He ions were applied to probe the tungsten simulation targets along the [001] direction. Deuterium atoms in the targets were detected using the nuclear reaction D($^3$He,p)$^4$He. A detector was located at 135$^{\circ}$ to detect the emitting protons from the reaction. The energy resolution of the detector was set to 16.5 keV. Differential cross sections of D($^3$He,p)$^4$He measured at 135$^{\circ}$ \cite{Wielunska2016} were used. The simulation temperature and the Debye temperature of the tungsten targets were 290 K and 377 K, respectively \cite{Walford1969}, which gives a 1D thermal vibration magnitude of tungsten atoms equal to 4 pm. The 1D thermal vibration magnitude of deuterium atoms was set to 14 pm \cite{Feldman1982}. The spread angle of emitting protons was set to 0$^{\circ}$, resulting in little effect of multiple scattering.
  
  \bigbreak
  
  Angular scans across the [001] axis were performed by gradually increasing the polar angle between the direction of the incident beam and the [001] direction. Results of angular scans can depend on the azimuthal angle between the incident beam and crystal axis as well. However, the azimuthal angle was not reported in the P-V experiments. Hence, for every polar angle, we accumulated NRA/C signals by varying the azimuthal angle from 0$^{\circ}$ to 359$^{\circ}$ with 1$^{\circ}$ as the increment. Thus, the simulation results at the negative polar angle region are a reflection of those at the positive region. The validity of this approach is presented in the Results section.   
  
  \bigbreak
  
\begin{figure}[h!]
  \centering
  \includegraphics[width=0.6\linewidth]{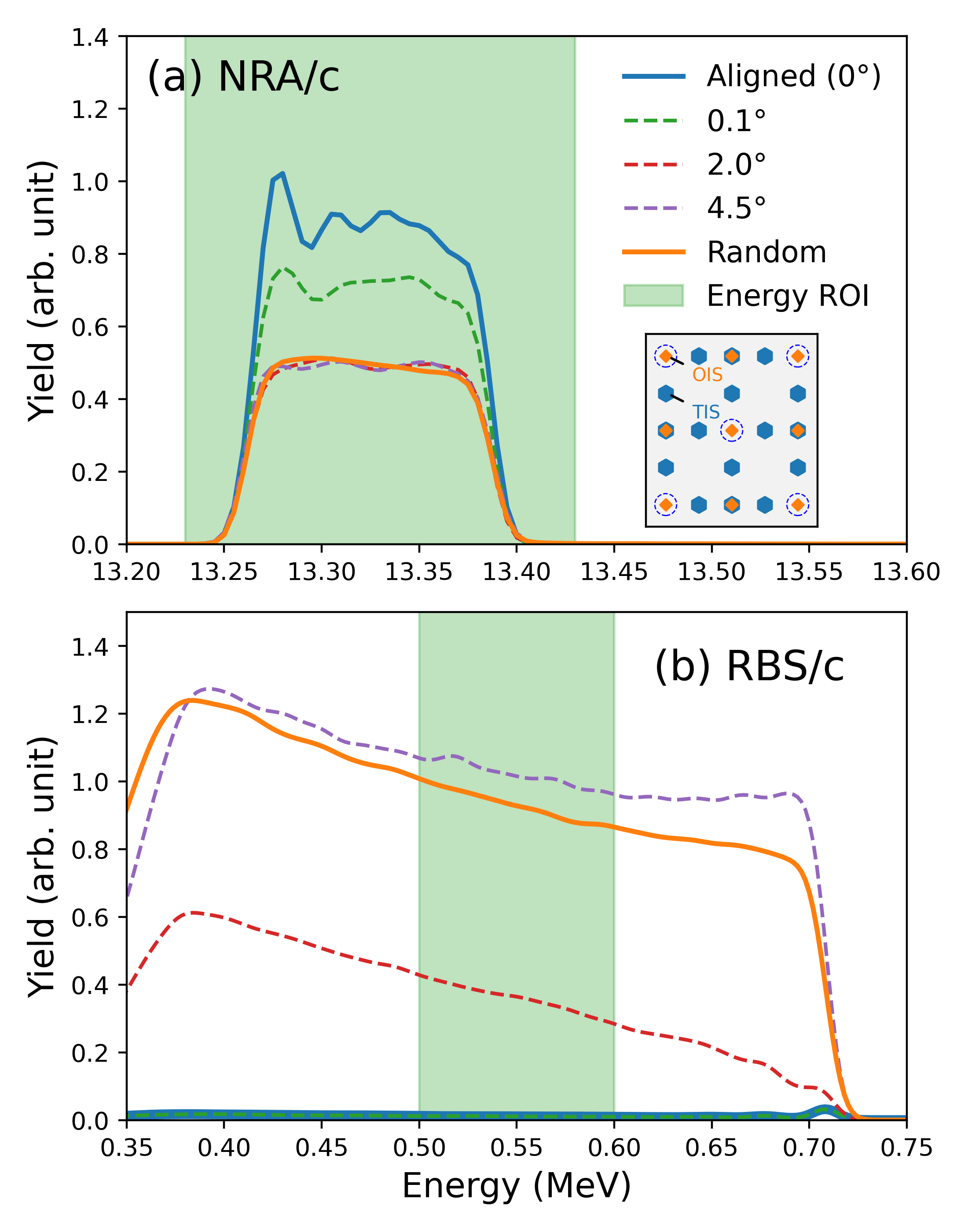}
  \caption{Example spectra obtained using 750 keV $^3$He ions on simulated targets along the [001] direction at different polar angles and at random configurations: (a) NRA/C spectra generated from targets with deuterium located at TIS, and (b) RBS spectra. The region of interest (ROI) for the angular scan calculations is represented by the colored region. The inset in (a) shows the locations of TIS (blue hexagon), OIS (orange square) and tungsten atoms (blue circle) viewed along the [001] direction. }
  \label{Fig:nra_rbs_spectra}
\end{figure}

   An example of NRA/C spectra of 750 keV $^3$He ions on tungsten containing 0.2 \% deuterium located at TIS are displayed in Fig. \ref{Fig:nra_rbs_spectra}a. The yield of spectra is highest when the incident beam is well aligned with the [001] axis (blue solid line), and decreases with a higher polar angle (dashed lines). We can observe an oscillation of the yield on the left side of the aligned spectrum. This is because the spatial distribution of $^3$He ions is not yet stable at the surface region. (Note the NRA/C signals generated from the surface region is at the low energy side of spectra.) In terms of angular scans, at each polar angle, the total yield of a whole spectrum was calculated, as indicated by the colored region of interest (ROI) in Fig. \ref{Fig:nra_rbs_spectra}a. Subsequently, the results were normalized by the total yield obtained from a random configuration (orange solid line), in which all the atoms in the simulation targets were randomly displaced.  
  
  \subsubsection{Setup of RBS/c simulations}
  
  The setup of RBS/c simulations is similar to that of NRA/C simulations. 750 keV $^3$He ions were applied to probe the tungsten simulation targets along the [001] direction. For RBS/c simulations, no deuterium is introduced to the simulation targets. $^3$He ions backscattered from tungsten atoms were detected by a detector located at 135$^{\circ}$ with an energy resolution of 16.5 keV. The 1D thermal vibration magnitude of tungsten atoms was 4 pm as in the NRA/C part. 
  
  \bigbreak
  
  Fig. \ref{Fig:nra_rbs_spectra}b shows example RBS/c spectra obtained at different polar angles and at the random configuration. The yield increases with a higher polar angle. For angular scans, the total yield in an energy range of 0.5 MeV to 0.6 MeV was calculated, and was subsequently normalized by that in the random configuration. 

\section{Results}

In this section, we show the variation of hydrogen lattice locations at different trapping conditions, and how this location variation affects the NRA/C signals. These results can be used as a comprehensive database for interpretation of NRA/C spectra with respect to their response to different types of deuterium trapping conditions.
 
 \bigbreak
 
 For each trapping condition, we compare the simulated NRA/C signals to that of P-V experiments, and assess the possibility of corresponding scenarios by taking into account the behavior of deuterium/hydrogen at room temperature in terms of solubility and binding energies. The analysis of trapping conditions starts with deuterium/hydrogen in perfect tungsten, followed by deuterium/hydrogen bonding to defects of interstitial and vacancy type.

 \subsection{Deuterium/Hydrogen in perfect tungsten}
 
\begin{figure}[h]
  \centering
  \includegraphics[width=1.0\linewidth]{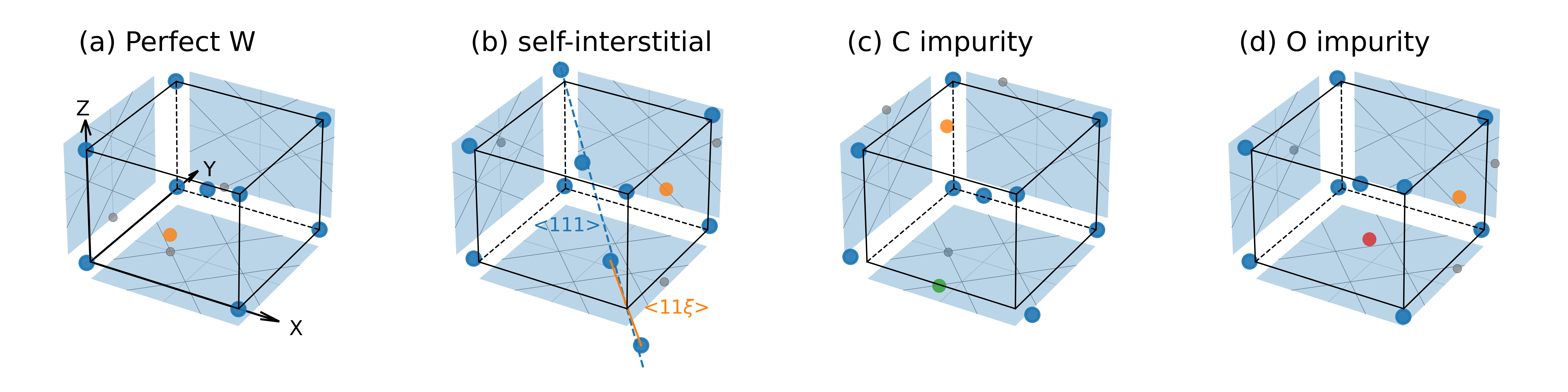}
  \caption{Locations of hydrogen (a) in perfect tungsten, (b) bonding to a $\langle 11\xi \rangle$ self-interstitial atom, (c) bonding to a carbon atom and (d) bonding to an oxygen atom as determined by DFT calculations. The tungsten, hydrogen, carbon and oxygen atoms are represented by the blue, orange, green and red circles, respectively. Projections of hydrogen atoms (gray) to three \{001\} planes are displayed on the blue planes, in which the intersection points of gray lines indicate the TIS. The $\langle 11\xi \rangle$ dumbbell is connected by the orange line.}
  \label{Fig:H_pos_int}
  \end{figure}

 The most stable site of a hydrogen atom in the perfect tungsten is at the TIS, as shown in Fig. \ref{Fig:H_pos_int}a, which is in agreement with other works \cite{Heinola2010, Fernandez2015}. The formation energy (or the solution energy) at the TIS, $E_f^{H,TIS}$, is 0.96 eV. After taking into account the ZPE correction, $E_f^{H,TIS}$ equals to 1.07 eV. (Three vibration frequencies of hydrogen at the TIS are 1571.6, 1568.8, 1166.5 cm$^{-1}$, respectively. The method of calculating ZPE is given in Sec.\ref{sec:DFT_method}.) A comparison of the formation energy of this work to that of other works and experiments can be found in Table.\ref{Tab:DFT_results}. As compared to the TIS, the hydrogen atom at the OIS has a higher formation energy ($E_f^{H,OIS} = 1.35$ eV). In fact, the vibration of hydrogen at the OIS has one real frequency and two imaginary frequencies (2521.2, $859.7i$ and $861.3i$ cm$^{-1}$). Thus, the hydrogen at the OTS is at a second order saddle point as reported in other works \cite{Fernandez2015}. In the following, when we calculate binding energies of hydrogen to defects, we consider that the hydrogen comes from a TIS far away from the defect. 
 
 \bigbreak
 
\begin{figure}[H]
  \centering
  \includegraphics[width=1.0\linewidth]{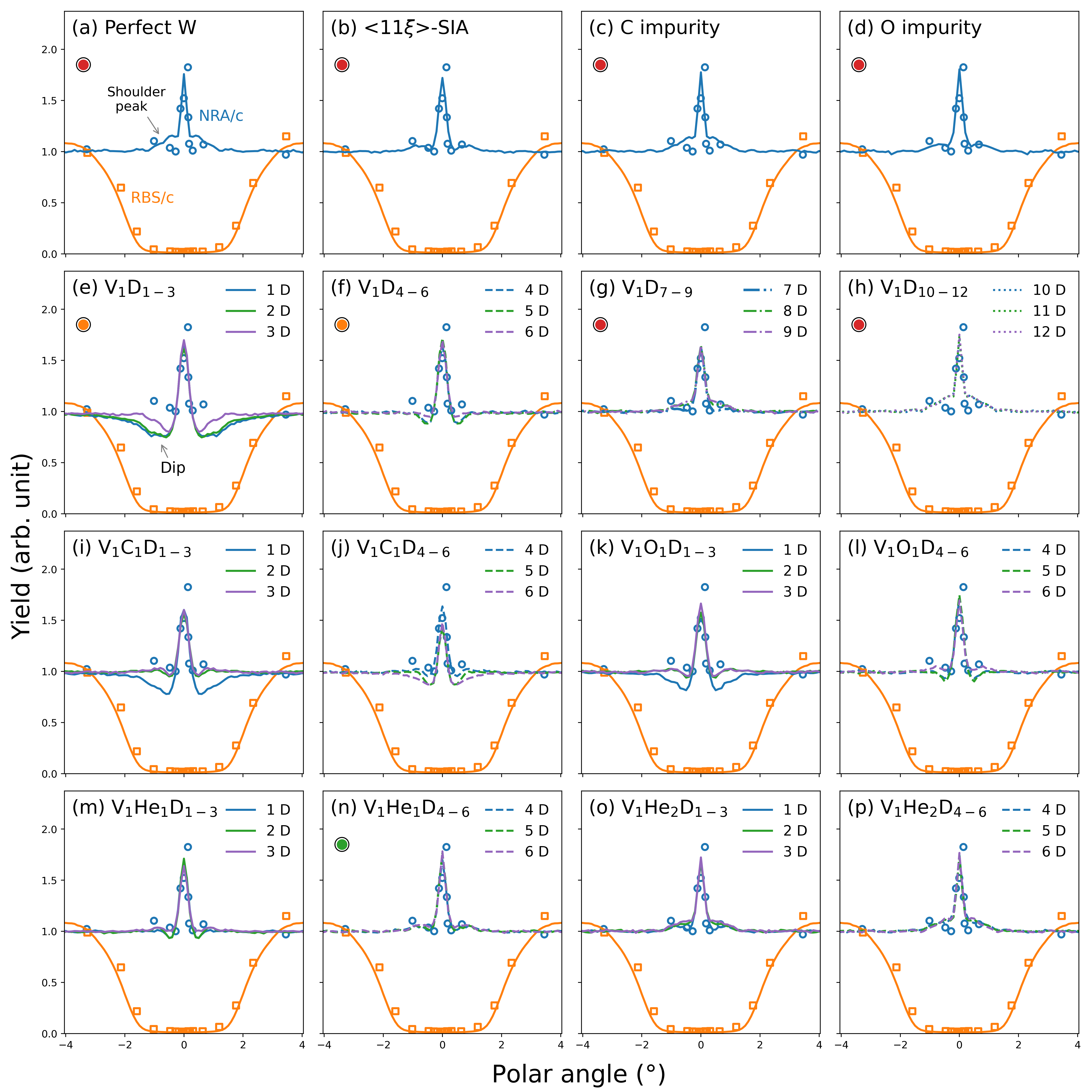}
  \caption{Comparison of experimental (dots) and simulated (lines) RBS/c and NRA/C angular scans through the [001] axis of tungsten, in which deuterium atoms are retained at different conditions: (a) retained in perfect tungsten, (b) bonding to a $\langle 11\xi \rangle$ self-interstitial atom, (c) bonding to a carbon atom, (d) bonding to an oxygen atom, (e-h) bonding to a vacancy with the number of deuterium atoms ranging from 1 to 12, (i-j) bonding to a vacancy containing a carbon atom (1 to 6 deuterium atoms), (k-l) bonding to a vacancy containing an oxygen atom (1 to 6 deuterium atoms), (m-n) bonding to a vacancy containing a helium atom (1 to 6 deuterium atoms), (o-p) bonding to a vacancy containing 2 helium atoms (1 to 6 deuterium atoms). The red, orange and green markers on the top left corner indicate a low possibility of corresponding condition (due to low solubility or low binding energy), significant difference between simulation and experiments for the corresponding condition and the most promising condition, respectively.(Experimental values are extracted from the P-V experiments \cite{Picraux1974}.)}
  \label{Fig:ang_scan}
  \end{figure}

Simulation results of angular scans through the [001] axis of tungsten are represented by lines in Fig. \ref{Fig:ang_scan}a, in which the results of P-V experiments are represented by markers. For RBS/c simulations, the results were obtained by using  perfect tungsten. The RBS/c simulation and experimental results share a good agreement, indicating the validity of the simulation method of angular scans. 
 
 \bigbreak
 
For NRA/C simulations, the signals were generated from a tungsten sample with deuterium atoms located at the TIS. The simulation results are close to those of P-V experiment. Both results feature a central peak at 0 $^{\circ}$, implying that there are deuterium atoms located at the center of the [001] channel. With a larger polar angle, we can observe two relatively small shoulder peaks on the two side of central peak. This should be attributed to deuterium atoms lying between the center of [001] channel and the row of tungsten atoms. However, the shoulder peaks in the P-V experiments are located at a larger angle than those in the simulations. Thus, the deuterium atoms in the P-V experiments are slightly displaced from the TIS. 
 
 \bigbreak
 
Although the simulation and experimental results seem to agree well except for some minor differences, the major contribution to experimental signals obtained at room temperature should not be due to the deuterium dissolved in tungsten because of the reasons discussed in the introduction (low solubility and high migration rate).  Most likely, the deuterium in the P-V experiments were attracted to defective structures. To explore this possibility, we consider next H atoms bonded to intrinsic defects.    
 
  \begin{table}[H]
  \begin{center}
  \caption{Comparison of the DFT results, including the lattice parameter of tungsten (Lat. para.), formation energy of a vacancy, $E_f^v$, formation energy of a $\langle 11\xi \rangle$ self-interstitial atom, $E_f^{i<11\xi>}$, formation energy of a hydrogen atom at the TIS, $E_f^{H,TIS}$, and at the OIS, $E_f^{H,OIS}$. (The unit of energy is in eV.)}
  \label{Tab:DFT_results}
  \begin{tabularx}{0.95\textwidth}{c|c|c|c}
  \toprule
  \multicolumn{1}{c}{} & \multicolumn{1}{c}{Present DFT} & \multicolumn{1}{c}{Other DFT} & \multicolumn{1}{c}{Experiments} \\
  \midrule
  
  \multicolumn{1}{c}{Lat. para. (Å)} & \multicolumn{1}{c}{3.172} & \multicolumn{1}{c}{3.172 \cite{Heinola2010a}} & \multicolumn{1}{c}{3.165 \cite{Parrish1960}} \\  
  
  \multicolumn{1}{c}{$E_f^v$} & \multicolumn{1}{c}{3.38} & \multicolumn{1}{c}{3.34 \cite{Heinola2010a}, 3.23 \cite{Becquart2010}} & \multicolumn{1}{c}{3.67 $\pm$ 0.2 \cite{Rasch1980}} \\
  
  \multicolumn{1}{c}{$E_f^{i<11\xi>}$} & \multicolumn{1}{c}{10.49} & \multicolumn{1}{c}{10.256$^b$ \cite{Ma2019}} & \multicolumn{1}{c}{} \\
  
  \multicolumn{1}{c}{$E_f^{H,TIS}$} & \multicolumn{1}{c}{0.96 (1.07)$^a$} & \multicolumn{1}{c}{0.95 \cite{Heinola2010}, 0.93 (1.04)$^a$ \cite{Fernandez2015}} & \multicolumn{1}{c}{1.04$\pm$0.17 \cite{Frauenfelder1969}} \\
  
  \multicolumn{1}{c}{$E_f^{H,OIS}$} & \multicolumn{1}{c}{1.35} & \multicolumn{1}{c}{1.34 \cite{Heinola2010}, 1.37 \cite{Fernandez2015}} & \multicolumn{1}{c}{} \\  
  
  \bottomrule
  \multicolumn{4}{l}{{\small $^a$ The value in the parenthesis is from calculations taking into account the ZPE correction.}} \\
  \multicolumn{4}{l}{{\small $^b$ The value is from calculations using $5 \times 5 \times 5$ supercells in ref.\cite{Ma2019}.}} \\
  \end{tabularx}
  \end{center}
  \end{table}
  
 \subsection{Deuterium/Hydrogen bonding to defects of interstitial type}
 
 We investigated the hydrogen bonding to a self-interstitial atom (SIA) in the $\langle 11\xi \rangle$ dumbbell configuration, which has been demonstrated to be the most stable configuration \cite{Ma2019}. This configuration is geometrically close to the $\langle 111 \rangle$ interstitial configuration with $\xi$ being an irrational number. In addition, since carbon and oxygen are major impurity atoms in tungsten \cite{Tanabe2014}, scenarios of hydrogen bonding to carbon and oxygen were also studied.
 
 \bigbreak
 
 The formation energy of the $\langle 11\xi \rangle$ SIA according to our calculation is 10.49 eV. The value of $\xi$ is close to 0.4, and the angle between the $\langle 11\xi \rangle$ and $\langle 111 \rangle$ direction is 19.4 $^{\circ}$. A comparison with that determined from other calculations is given in Table.\ref{Tab:DFT_results}. To study the position of a hydrogen bonding to the $\langle 11\xi \rangle$ SIA, a hydrogen atom was initially placed in the vicinity of $\langle 11\xi \rangle$ SIA. After relaxation, the hydrogen atom is located at a position close to the TIS as shown in Fig. \ref{Fig:H_pos_int}b, which is displaced by 0.22 Å from the TIS towards OIS. (The distance between the TIS and OIS is 0.793 Å.) The distance of the hydrogen atom to the center of $\langle 11\xi \rangle$ SIA is 2.60 Å. The binding energy of hydrogen to the $\langle 11\xi \rangle$ SIA is 0.23 eV (0.28 eV after taking into account the ZPE correction). We have also explored the scenarios of placing a hydrogen atom at the vicinity of $\langle 111 \rangle$ SIA. After relaxation, the $\langle 111 \rangle$ SIA is distorted and transforms to $\langle 11\xi \rangle$ SIA. The hydrogen atom is still close to the TIS. Distortion of $\langle 111 \rangle$ SIA by hydrogen atoms has also been reported in other studies \cite{Heinola2010a}.
 
 \bigbreak
 
 In terms of the impurity atoms, carbon prefers to occupy the OIS in a perfect tungsten. If we put a carbon atom at a TIS, it moves to an OIS after position relaxation. On the contrary, the stable position for oxygen is the TIS. These results are in agreement with those reported in other works \cite{LiuYL2010, Ou2012, KongXS2013}. As in the case of <111> SIA, a hydrogen atom was initially placed in the vicinity of a carbon atom. After position relaxation, the hydrogen atom is almost located at the TIS (displaced by 0.03 Å from the TIS to OIS). The distance between the hydrogen and carbon atom is 3.60 Å. The binding energy in this case is very weak, 0.05 eV (0.06 eV with the ZPE correction). The hydrogen bonding to an oxygen atom is displaced by 0.16 Å from the TIS towards OIS. The distance and the binding energy between the hydrogen and oxygen atom are 2.24 Å and 0.24 eV (0.26 eV with the ZPE correction), respectively. Fig. \ref{Fig:H_pos_int}c and Fig. \ref{Fig:H_pos_int}d show the positions of hydrogen bonding to the carbon and oxygen atom, respectively.

 \bigbreak
 
  NRA/C results for deuterium bonding to the $\langle 11\xi \rangle$ SIA, carbon and oxygen atom are shown in Fig. \ref{Fig:ang_scan}b, Fig. \ref{Fig:ang_scan}c and Fig. \ref{Fig:ang_scan}d, respectively. Since the positions of deuterium in these cases are close to the TIS, the simulation results appear to have good agreement with the experimental ones, especially for the cases of $\langle 11\xi \rangle$ SIAs and oxygen. However, the binding energy of deuterium/hydrogen to SIAs is low. It has been shown that the detrapping temperature of hydrogen bonding to a $\langle 111 \rangle$ SIA is lower than room temperature (below 200 K) \cite{Heinola2010a}. Also the SIA itself is extremely mobile even at cryogenic temperatures, and it is very unlikely that isolated SIAs would be present in experimental samples \cite{Swi17} (indeed, field ion microscopy experiments have observed vacancies, but not interstitials in tungsten \cite{Cur81,Wei81}). Similarly, we can expect a low detrapping temperature for hydrogen bonding to a carbon and oxygen atom. In fact, the binding energy of hydrogen to these defects depends on the distance, and remains at a relatively small level \cite{KongXS2013, Backer2017}. Thus, we can suggest that the signals in the P-V experiments are mainly contributed by deuterium trapped by other types of defects.

 \subsection{Deuterium/Hydrogen bonding to vacancies} \label{sec:H2Vac}
 
The energy of deuterium atoms (30 keV) in the P-V experiments is high enough to create vacancies. Considering that the maximum concentration of vacancy is low (0.065 \%) as presented in Fig. \ref{Fig:srim_profile} and the migration energy of a vacancy is high (1.71 eV from ref.\cite{Heinola2010a}), it is fair to assume that at room temperature the vacancies in the experimental samples will be present mainly as monovacancies \cite{Granberg2021}. Hence, in the present study we investigated the scenarios with multiple hydrogen atoms trapped in a mono-vacancy.
 
 \bigbreak
 
\begin{figure}[H]
  \centering
  \includegraphics[width=1.0\linewidth]{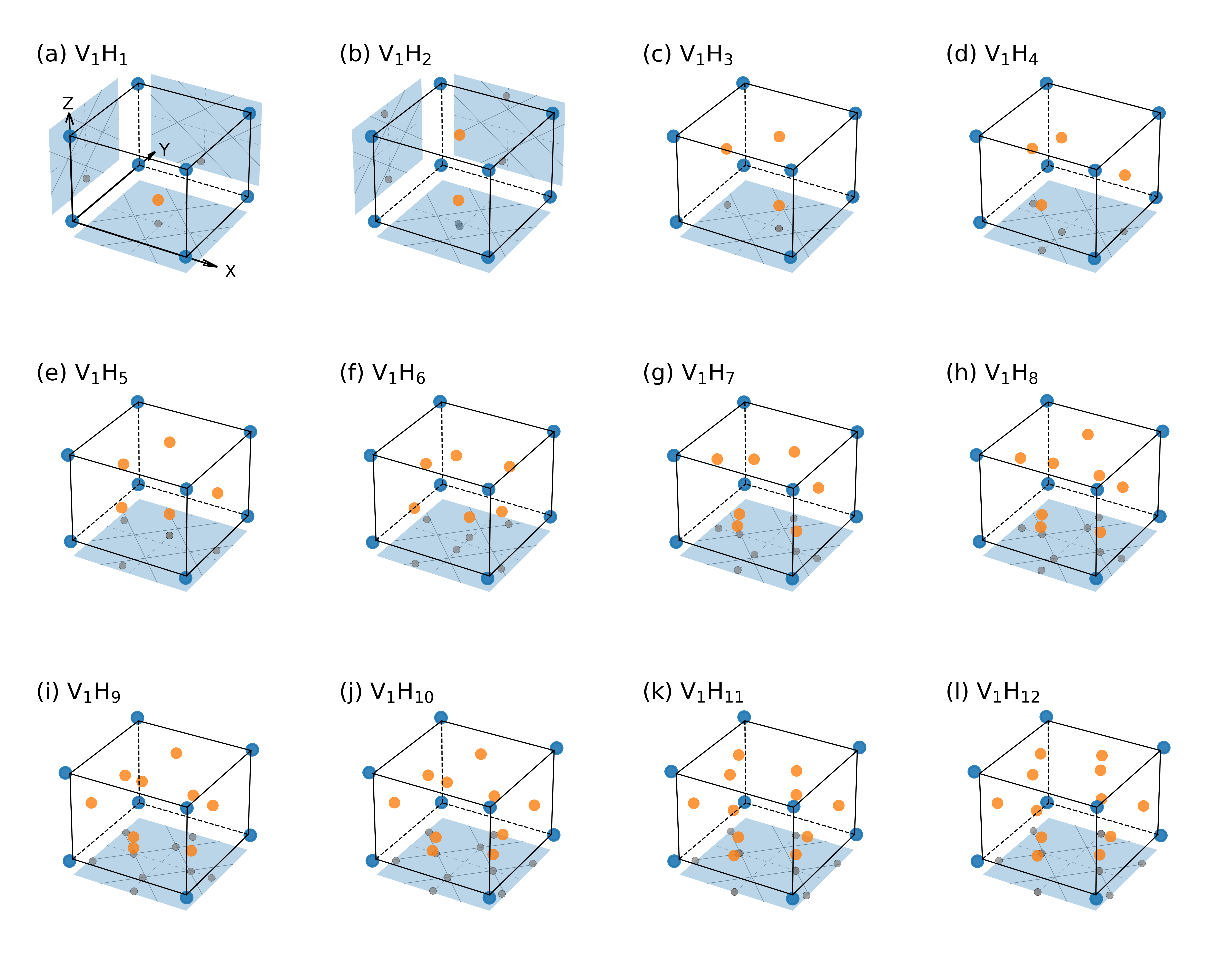}
  \caption{Positions of hydrogen atoms (orange) in a vacancy: from (a) to (l) the number of hydrogen atoms is increased from 1 to 12. Other notations are same to those in Fig. \ref{Fig:H_pos_int}. For simplicity, the projections to the (100) and (010) planes are not shown for cases with more than 2 hydrogen atoms.)}
  \label{Fig:H_pos_vac}
 \end{figure}

 The formation energy of vacancy according to our calculations is 3.38 eV. A comparison of the formation energy for this work to that of other works and experiments can be found in Table \ref{Tab:DFT_results}. Hydrogen atoms were introduced to a vacancy by two methods. In the first method, the initial positions of hydrogen atoms were randomly selected inside a vacancy. In the second method, we introduced the hydrogen atoms following the stable structures proposed by Ohsawa \textit{et al} \cite{Ohsawa2010}. After positions relaxation, we took the structure with a smaller energy obtained from the two methods. It turns out that the distributions of hydrogen atoms with minimum energies, as shown in Fig. \ref{Fig:H_pos_vac}, are similar to those in the work of Ohsawa \textit{et al} \cite{Ohsawa2010}.
 
 \bigbreak
 
\begin{figure}[H]
  \centering
  \includegraphics[width=0.6\linewidth]{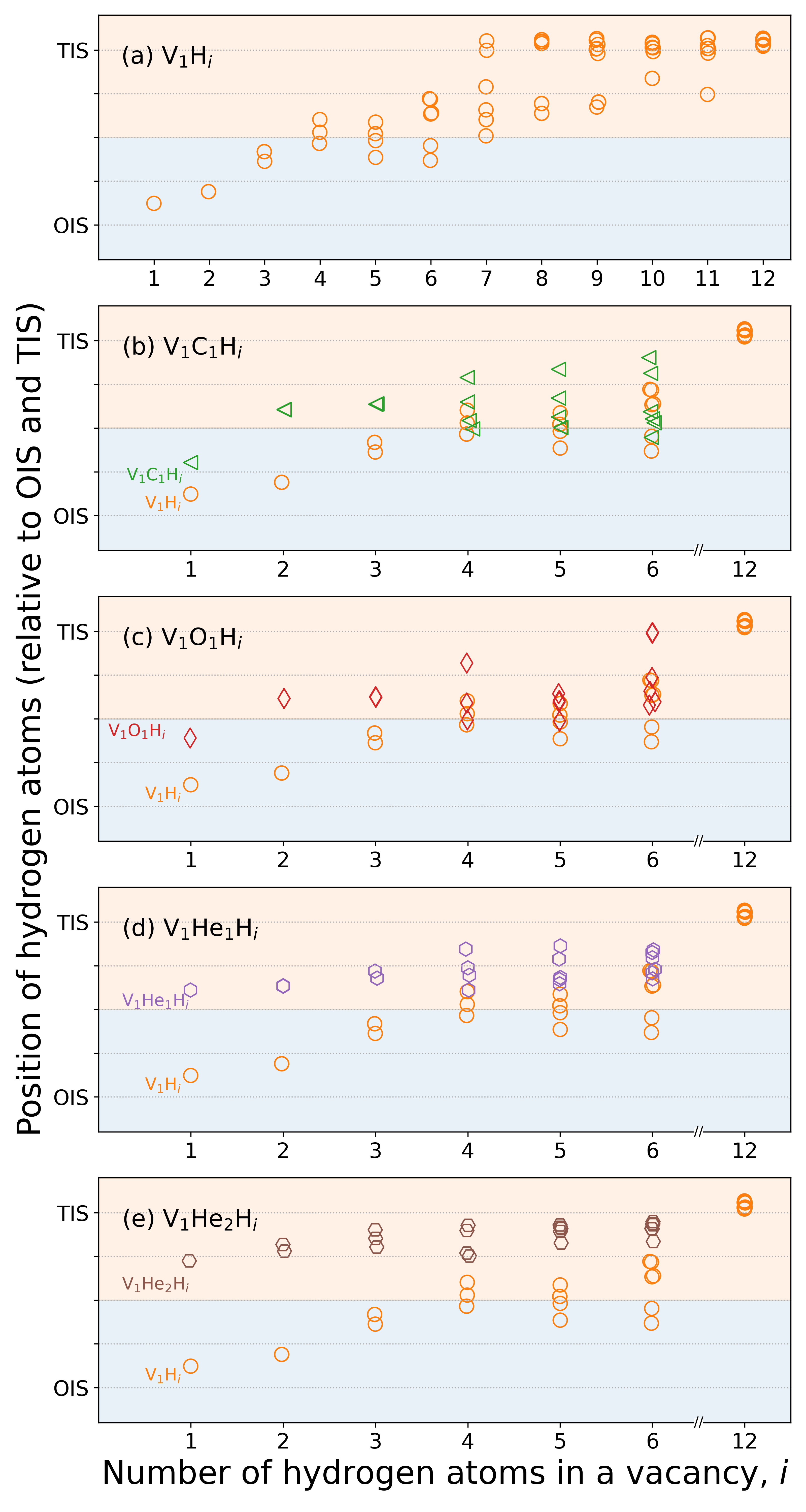}
  \caption{Positions of hydrogen atoms, relative to the OIS and TIS, as a function of the number of hydrogen atoms in a vacancy or vacancy complex. The hydrogen atoms are bonding to (a) a vacancy (orange) and a vacancy complex containing (b) a carbon (green), (c) an oxygen (red), (d) a helium atom (purple) and (e) two helium atoms (brown), respectively. (Note that some symbols are overlapping)}
  \label{Fig:pos_evolve_every}
 \end{figure}

 Fig. \ref{Fig:pos_evolve_every}a illustrates the positions of hydrogen atoms bonding to a vacancy with different filling levels, in terms of the relative distance to the OIS and TIS. The notation V$_1$H$_i$ represents $i$ hydrogen atoms bonding to one vacancy. When there is only one hydrogen atom bonding to the vacancy, the hydrogen atom occupies a position close to the OIS. It is slightly displaced by 0.10 Å from the OIS towards the TIS, and has a distance of 0.32 Å with the closest vacancy surface (a (001) plane). When there are two hydrogen atoms, the hydrogen atoms occupy positions nearby the OIS, but are located at two opposite planes instead of being at the same plane. This distribution can be attributed to a generally repulsive nature of hydrogen-hydrogen interaction at short distances inside tungsten \cite{Hou2019}. With a higher number of hydrogen atoms, hydrogen atoms start to occupy remaining empty planes surrounding the vacancy. Each plane holds one hydrogen atom until there is a total number of 6 hydrogen atoms in the vacancy. Noticeably, the stable position of hydrogen atom is gradually shifted from the OIS towards the direction of TIS. When the trapping number is 6, the hydrogen atoms are distributed around the midpoint of the OIS and TIS.
 
 \bigbreak
 
\begin{figure}[H]
  \centering
  \includegraphics[width=0.6\linewidth]{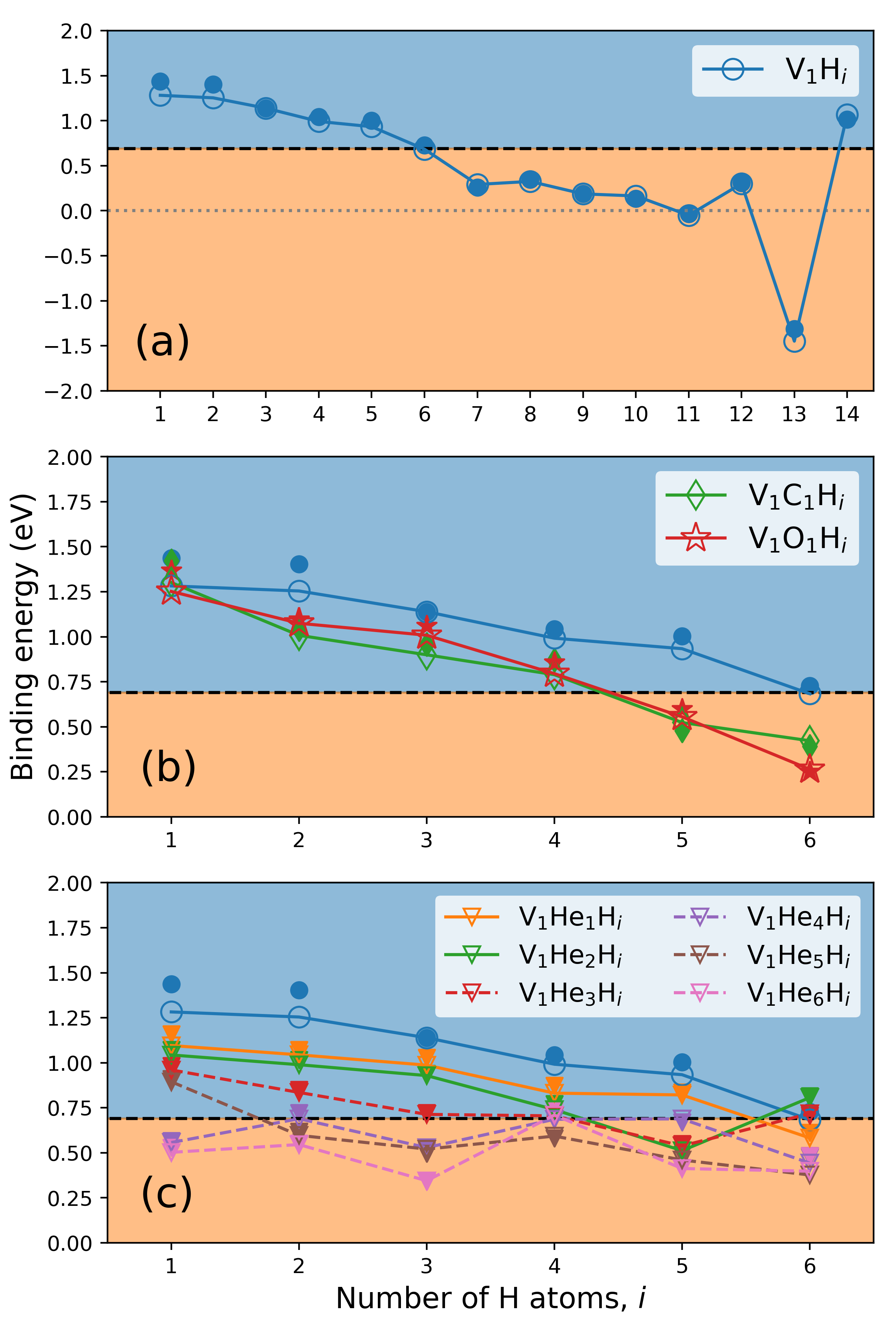}
  \caption{Binding energies of a hydrogen atom to vacancy complexes: (a) a vacancy with $(i-1)$ hydrogen atoms, (b) a vacancy containing a carbon (green) or oxygen (red) atom and $(i-1)$ hydrogen atoms, (c) a vacancy containing multiple helium and $(i-1)$ hydrogen atoms. The binding energies without and with the ZPE correction are represented by hollow and solid markers, respectively. The black dashed line represents an estimated threshold energy for the detrapping of hydrogen at room temperature.}
  \label{Fig:bind_ener}
  \end{figure}

 When the total number of hydrogen atoms in the vacancy is larger than 6, some planes surrounding the vacancy start to accommodate two hydrogen atoms, in which case the two hydrogen atoms are located at positions close to the TIS. Thus, when 12 hydrogen atoms are trapped by the vacancy, each plane accommodates two hydrogen atoms, and all the hydrogen atoms are basically positioned at the TIS. The binding energy of hydrogen to the vacancy gradually diminishes with a higher number of hydrogen atoms as given in Fig. \ref{Fig:bind_ener}a, in which the binding energy denoted by VH$_i$ represents the binding energy of a hydrogen atom located at a TIS far away from the vacancy to a vacancy containing $i-1$ hydrogen atoms (the binding energies without and with the ZPE correction are represented by the blue hollow and filled circles, respectively). The results indicate that a vacancy can hold up to 12 hydrogen atoms. In case that 14 hydrogen atoms are introduced to a vacancy, a hydrogen molecule is formed, resulting in an increase of binding energy. These features are in line with those reported in other works \cite{Ohsawa2010, Fernandez2015}

  \bigbreak
  
  Fig. \ref{Fig:ang_scan}(e-h) show NRA/C results of a vacancy containing up to 12 deuterium atoms, which exhibit distinct features with different filling levels of deuterium.  For the cases of V$_1$D$_{1-3}$, the NRA/C results exhibit significant differences with that of deuterium in perfect tungsten, as shown in Fig. \ref{Fig:ang_scan}a as well as to the P-V experiments. This is because, in the cases of a vacancy containing 1 to 3 deuterium atoms, the deuterium atoms are close to the OIS, and 1/3 of OISs are covered by tungsten atoms viewed along the [001] direction. Thus, instead of shoulder peaks shown in Fig. \ref{Fig:ang_scan}a (in which deuterium atoms are at the TIS), we can observe dips at the two sides of the central peak. More specifically, for the case of V$_1$D$_3$, the dips become narrower because the deuterium atoms are displaced from the OIS by a larger distance.
  
  \bigbreak
  
  By further increasing the number of deuterium atoms in the vacancy, it can be clearly observed that the signals of dips fade as shown in Fig. \ref{Fig:ang_scan}f. For the case of V$_1$D$_6$, the dips almost disappear, and the NRA/C signals feature only a central peak. When the total number of deuterium is larger than 6, the signals of shoulder peaks are gradually built up as shown in Fig. \ref{Fig:ang_scan}g. Finally, the NRA/C signals become stable for the cases of a vacancy containing 10 to 12 deuterium atoms as shown in Fig. \ref{Fig:ang_scan}h, and are similar to that of deuterium in perfect tungsten shown in Fig. \ref{Fig:ang_scan}a.
  
  \bigbreak
  
Comparing the NRA/C simulation and the P-V experiments, the spectra obtained with vacancies containing more than 6 deuterium atoms are consistent with experiments. However, it has been suggested that the filling level of deuterium/hydrogen in a vacancy is only 6 at room temperature according to the calculation performed by Fernandez \textit{et al.} \cite{Fernandez2015}. By assuming a threshold binding energy to be 0.69 eV (which is the binding energy of V$_1$H$_6$ according to the results given by Fernandez \textit{et al.} \cite{Fernandez2015}), we can divide the trapping of deuterium/hydrogen in a vacancy to two categories, as indicated by the blue and orange regions in Fig. \ref{Fig:bind_ener}. Clearly, the binding energies of a vacancy containing 7 to 12 deuterium/hydrogen atoms are smaller than the threshold. Hence, it is less likely that in the P-V experiments vacancies were containing more than 6 deuterium atoms.         
  
  \bigbreak
  
  At this stage, we have excluded a number of trapping conditions based on the evaluation of binding energy. Only remaining scenarios are the cases of a vacancy containing 1 to 6 deuterium atoms. Yet, the NRA/C signals of the remaining scenarios are clearly different with the P-V experiments. Note that this difference does not mean that the deuterium atoms are not trapped by the vacancies in the experiments. Instead, it indicates that there must be some additional factors affecting the positions of deuterium, for example, impurity atoms in the vacancy.

 \subsection{Deuterium/Hydrogen bonding to vacancies with carbon or oxygen}
 
 Carbon and oxygen show high binding energies to a vacancy, 2.16 eV and 3.26 eV according to our calculations, respectively. Inside a vacancy, both atoms prefer to occupy a position close to the OIS, which is in agreement with that reported in other works \cite{KongXS2013}. We introduced up to 6 hydrogen atoms in vacancies containing a carbon or oxygen atom. The introduction method is the same to that presented in previous section. The corresponding binding energies are given in Fig. \ref{Fig:bind_ener}b, in which the notation V$_1$C$_1$H$_i$ (V$_1$O$_1$H$_i$) represents a vacancy containing $i$ hydrogen atoms and a carbon (oxygen) atom.
 
 \bigbreak 
 
\begin{figure}[H]
  \centering
  \includegraphics[width=1.0\linewidth]{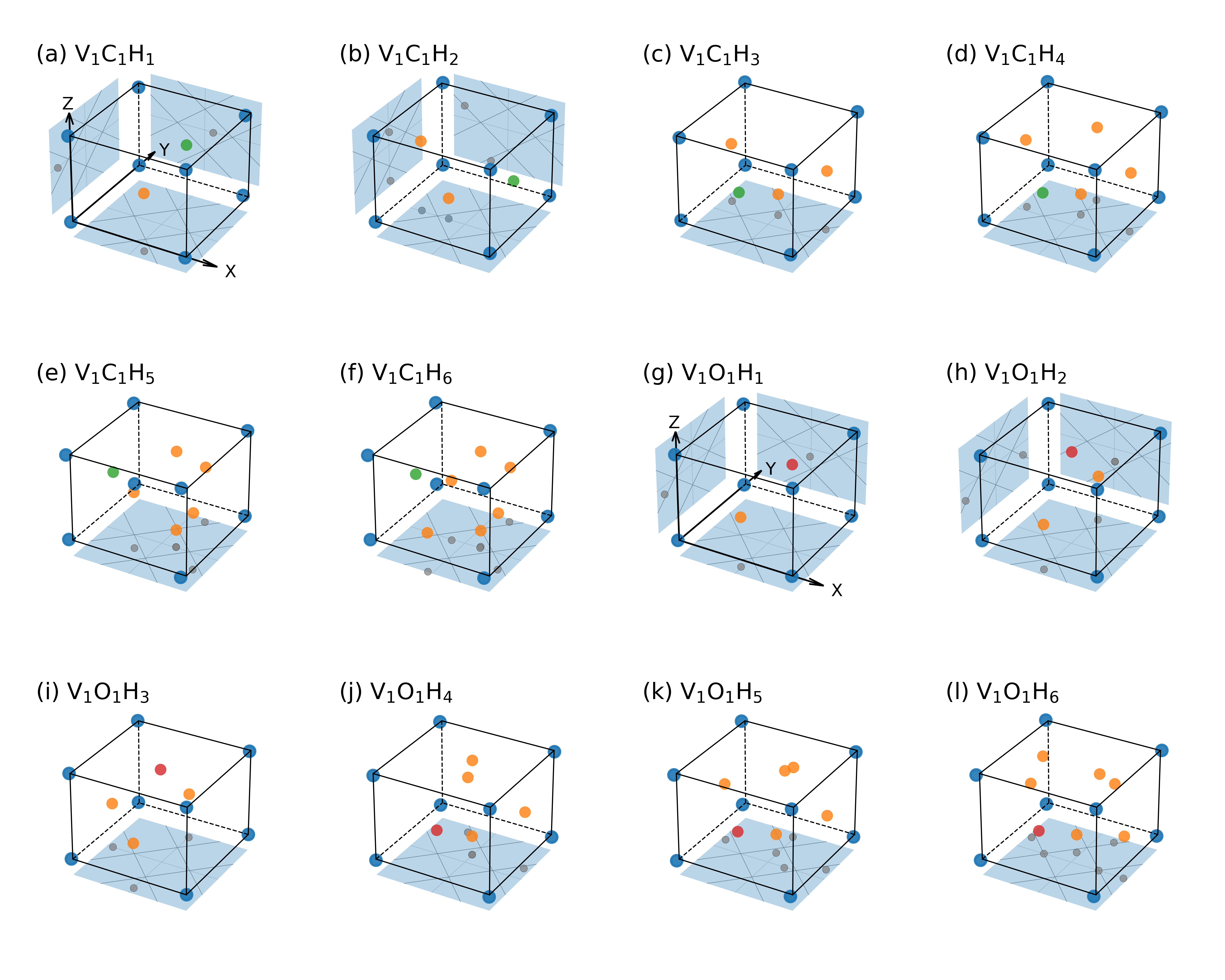}
  \caption{Positions of hydrogen atoms (orange) in a vacancy containing a carbon atom (a-f) and containing an oxygen atom (g-l). The carbon and oxygen atoms are represented by the green and red circles, respectively. Other notations are the same to those in Fig. \ref{Fig:H_pos_int}. (The projections to the (100) and (010) planes are not shown for vacancies with more than 2 hydrogen atoms for the sake of simplicity.)}
  \label{Fig:H_pos_vac_CO}
 \end{figure}

The positions of hydrogen atoms in a vacancy containing a carbon atom are shown in Fig. \ref{Fig:H_pos_vac_CO}(a-f). Fig. \ref{Fig:pos_evolve_every}b illustrates the relative positions of hydrogen atoms between the OIS and TIS. The corresponding NRA/C results are shown in Fig. \ref{Fig:ang_scan}(i-j). The insertion of a carbon atom into a vacancy indeed affects the stable positions of hydrogen atoms. For the case of V$_1$C$_1$H$_1$, the hydrogen atom is still close to the OIS, but the distance to the OIS becomes larger compared to the case without carbon. This displacement is reflected from the narrower NRA/C dips in Fig. \ref{Fig:ang_scan}i compared to that in Fig. \ref{Fig:ang_scan}e. When the number of hydrogen atoms is increased to 3, the hydrogen atoms are further displaced towards the TIS. The corresponding NRA/C results are featured with small shoulder peaks as shown in Fig. \ref{Fig:ang_scan}i. 

\bigbreak 
 
However, when the number of hydrogen atom is increased from 4 up to 6, the configuration of hydrogen atoms become spread: some hydrogen atoms get closer to the TIS, but some are slightly displaced backwards towards the direction of OIS as shown in Fig. \ref{Fig:pos_evolve_every}b. In addition, for the cases of V$_1$C$_1$H$_5$ and V$_1$C$_1$H$_6$ hydrogen atoms, one hydrogen atom is located between the carbon atom and vacancy center instead of the vacancy surface. As shown in Fig. \ref{Fig:ang_scan}j, due to the spread of deuterium/hydrogen atoms, from V$_1$C$_1$H$_4$ to V$_1$C$_1$H$_6$, the central peak of NRA/C slightly diminishes accompanied by a recurrence of dips.  
 
 \bigbreak
 
  The positions of hydrogen atoms in a vacancy containing an oxygen atoms are shown in Fig. \ref{Fig:H_pos_vac_CO}(g-l). Fig. \ref{Fig:pos_evolve_every}c illustrates the relative positions of hydrogen atoms between the OIS and TIS. The corresponding NRA/C results are shown in Fig. \ref{Fig:ang_scan}(k-l).The oxygen atom affects the positions of hydrogen atoms in a similar way as the carbon atom. From V$_1$O$_1$H$_1$ to V$_1$O$_1$H$_3$, the hydrogen atoms are shifted towards the TIS. As a result, we can observe a transformation of dips to small shoulder peaks in the NRA/C results (see Fig. \ref{Fig:ang_scan}k). When there are 4 and 5 hydrogen atoms, the positions of hydrogen atoms are spread around the midpoint of OIS and TIS. For the case of 6 hydrogen atoms, two hydrogen atoms are positioned on a same vacancy surface and occupies TISs. Thus, the corresponding NRA/C results, as shown in Fig. \ref{Fig:ang_scan}l, exhibit shoulder peaks instead of dips as in the case of V$_1$C$_1$H$_6$ (see Fig. \ref{Fig:ang_scan}j). 
 
 \bigbreak
 
 In general, after considering the effect of carbon and oxygen atoms, the agreement between the NRA/C simulation and experiments is improved. Nonetheless, the central peaks in the cases of V$_1$C$_1$D$_i$ are clearly lower than those in the experiments. For the cases of V$_1$O$_1$D$_i$, the NRA/C result generated from a vacancy containing 6 deuterium atoms shows the best agreement with the experiment. However, the binding energy for this case is lower than the threshold binding energy as shown in Fig. \ref{Fig:bind_ener}b, implying a possible detrapping of deuterium/hydrogen atoms at room temperature. Again, we need to mention that as long as the binding energies are high, the corresponding scenarios are possible. Nonetheless, in order to get a better agreement between NRA/C simulations and experiments, we propose to investigate the effect of another element, i.e., helium.

 \subsection{Deuterium/Hydrogen bonding to vacancies with helium}
 
 A helium atom in a perfect tungsten prefers to occupy the TIS with a formation energy of 6.14 eV (6.23 eV with the ZPE correction), in agreement with other DFT calculations \cite{Becquart2009}. It has also been shown that a single tungsten vacancy  can accommodate at least 9 helium atoms \cite{Takayama2013}. Here, we investigated the scenarios of a vacancy containing from 1 to 6 helium atoms. Contrary to hydrogen, our calculations show that the preferential position of a helium atom in a vacancy is at its center, which is in line with other studies \cite{ZhouHB2016}. 
 
 \bigbreak
 
 Increasing the number of helium atoms in a vacancy did not change the trend: the helium atoms still gathered close to the vacancy center. The configurations of multiple helium atoms in a vacancy exhibit high symmetry features. From 2 to 6 helium atoms, the helium atoms are organized in a shape of <111>-oriented dumbbell, equilateral triangle (with a side length of 1.53 Å), regular tetrahedron (with a side length of 1.56 Å), triangular bipyramid and regular octahedron (with a side length of 1.57 Å), respectively. Here, we show the cases of a vacancy containing up to 2 helium atoms and multiple hydrogen atoms. The cases of a vacancy containing 3 to 6 helium atoms are presented in the supplementary materials. The notation V$_1$He$_n$H$_i$, represents a vacancy containing $n$ helium atoms and $i$ hydrogen atoms.
 
 \bigbreak
 
 We found that it is prone for a vacancy complex containing both helium and hydrogen atoms to form a meta-stable structure. Thus, when we introduce hydrogen atoms to the vacancy complex, in addition to the method used in Sec.\ref{sec:H2Vac}, we assume that, in general, the binding energy of a hydrogen atom to the vacancy complex decreases with a higher number of hydrogen and helium atom. For cases against this assumption, we perform additional calculations by adjusting initial positions of atoms in the vacancy. 
 
 \bigbreak
 
\begin{figure}[H]
  \centering
  \includegraphics[width=1.0\linewidth]{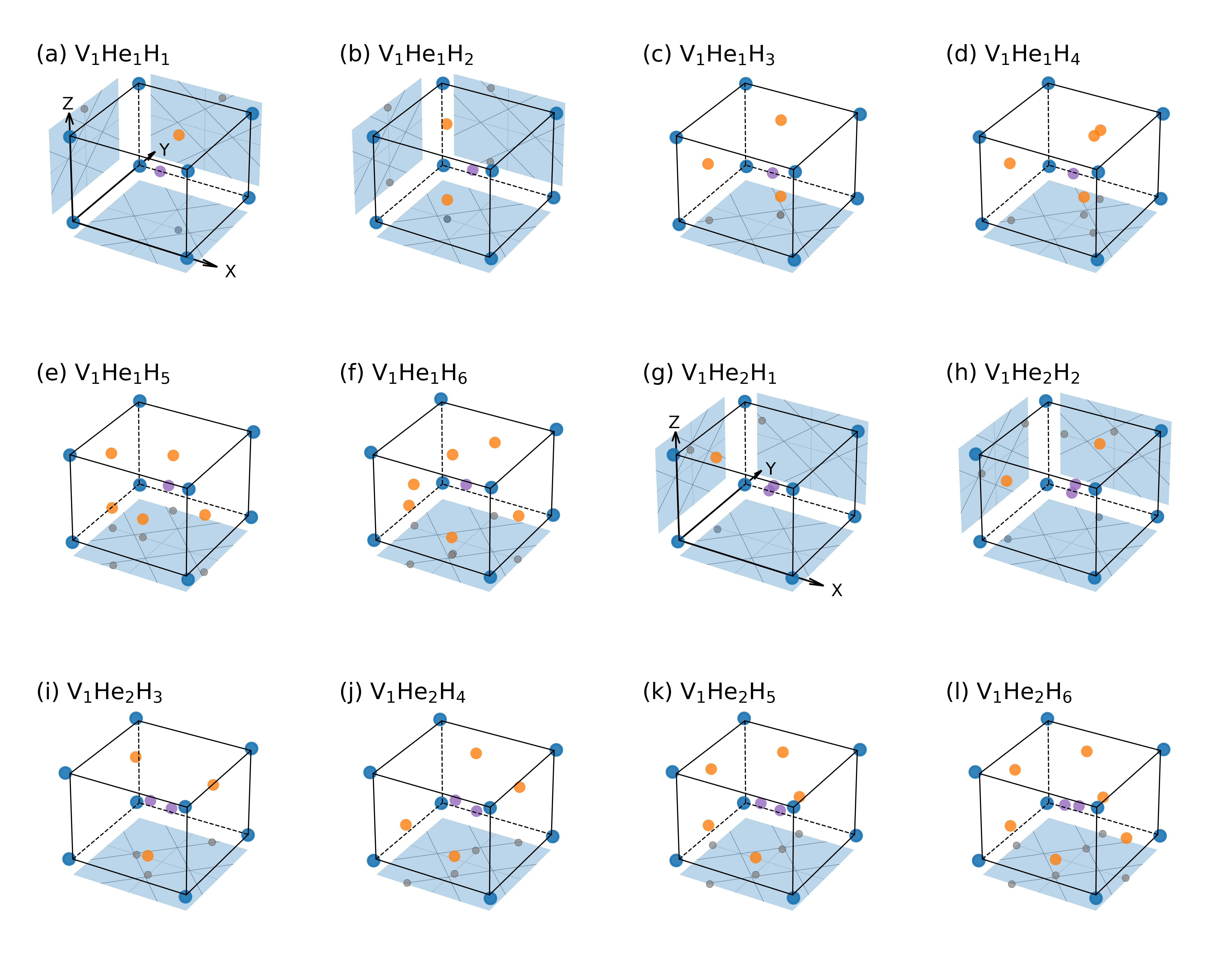}
  \caption{Positions of hydrogen atoms (orange) in a vacancy containing one helium atom (a-f) and containing two helium atoms (g-l). The helium atoms are represented by the purple circles. Other notations are the same to those in Fig. \ref{Fig:H_pos_int}. (The projections to the (100) and (010) planes are not shown for vacancies with more than 2 hydrogen atoms for the sake of simplicity.)}
  \label{Fig:H_pos_vac_He12}
 \end{figure}

 Stable positions of hydrogen atoms for the cases of V$_1$He$_1$H$_i$ and V$_1$He$_2$H$_i$ are given in Fig. \ref{Fig:H_pos_vac_He12}(a-f) and Fig. \ref{Fig:H_pos_vac_He12}(g-l), respectively. Fig. \ref{Fig:pos_evolve_every}d and Fig. \ref{Fig:pos_evolve_every}e illustrate the relative positions of hydrogen atoms between the OIS and TIS for the cases with one and two helium atoms, respectively. The corresponding binding energies are shown in Fig. \ref{Fig:bind_ener}c, in which the binding energies for vacancies containing 3 to 6 helium atoms are also displayed. The binding energies of V$_1$He$_1$H$_i$ and V$_1$He$_2$H$_i$ follow the assumption mentioned before (the binding energy decreases with more hydrogen and helium atoms) except the case of V$_1$He$_2$H$_6$. One plausible reason is that the configuration of hydrogen atoms in this particular case possesses a higher symmetry compared to that of V$_1$He$_2$H$_5$.
 
 \bigbreak
 
 For the case of V$_1$He$_1$H$_1$, the hydrogen atom is closer to the TIS (displaced by 0.48 Å from the OIS to TIS). This is in line with other DFT calculations showing hydrogen atoms prefer to occupy TISs instead of OISs \cite{Jiang2010}. By increasing the number of hydrogen atoms to 3, we can observe a shorter distance between the hydrogen atoms and TIS. The corresponding NRA/C results are given in Fig. \ref{Fig:ang_scan}m, in which there is a significant shrink of dips compared to the cases without helium atoms (see Fig. \ref{Fig:ang_scan}e). For the cases of V$_1$He$_1$H$_{4-6}$, the stable positions of hydrogen atoms are further adjusted towards the TIS. The corresponding NRA/C results are featured with small shoulder peaks resembling the experimental ones as shown in Fig. \ref{Fig:ang_scan}n.
 
 \bigbreak
 
 After introducing the second helium atom to above vacancies, the distance of hydrogen atoms and the TIS becomes smaller. Noticeably, the relative positions of hydrogen atoms are modified in some cases. For the case of V$_1$He$_2$H$_2$, the configuration of 2 hydrogen atoms located at two adjacent vacancy surfaces as shown in Fig. \ref{Fig:H_pos_vac_He12}h has a lower energy than a metastable configuration in which 2 hydrogen atoms are located at two opposite surfaces. For the case of V$_1$He$_2$H$_4$, the relative positions of hydrogen atoms are also re-distributed as shown in Fig. \ref{Fig:H_pos_vac_He12}j (the hydrogen atoms are at two opposite (001) planes and two opposite (010) planes) compared to that in Fig. \ref{Fig:H_pos_vac_He12}d (two opposite (001) planes and adjacent (100) and (010) planes). In terms of the corresponding NRA/C results as shown in Fig. \ref{Fig:ang_scan}o and Fig. \ref{Fig:ang_scan}p, the signals of shoulder peaks are enhanced as compared to those in the cases of one helium atom. For vacancies containing more helium atoms (up to 6 helium atoms), the hydrogen atoms are still near the TIS, which are presented in the supplementary materials.  
 
 \bigbreak
 
 \begin{figure}[H]
  \centering
  \includegraphics[width=0.7\linewidth]{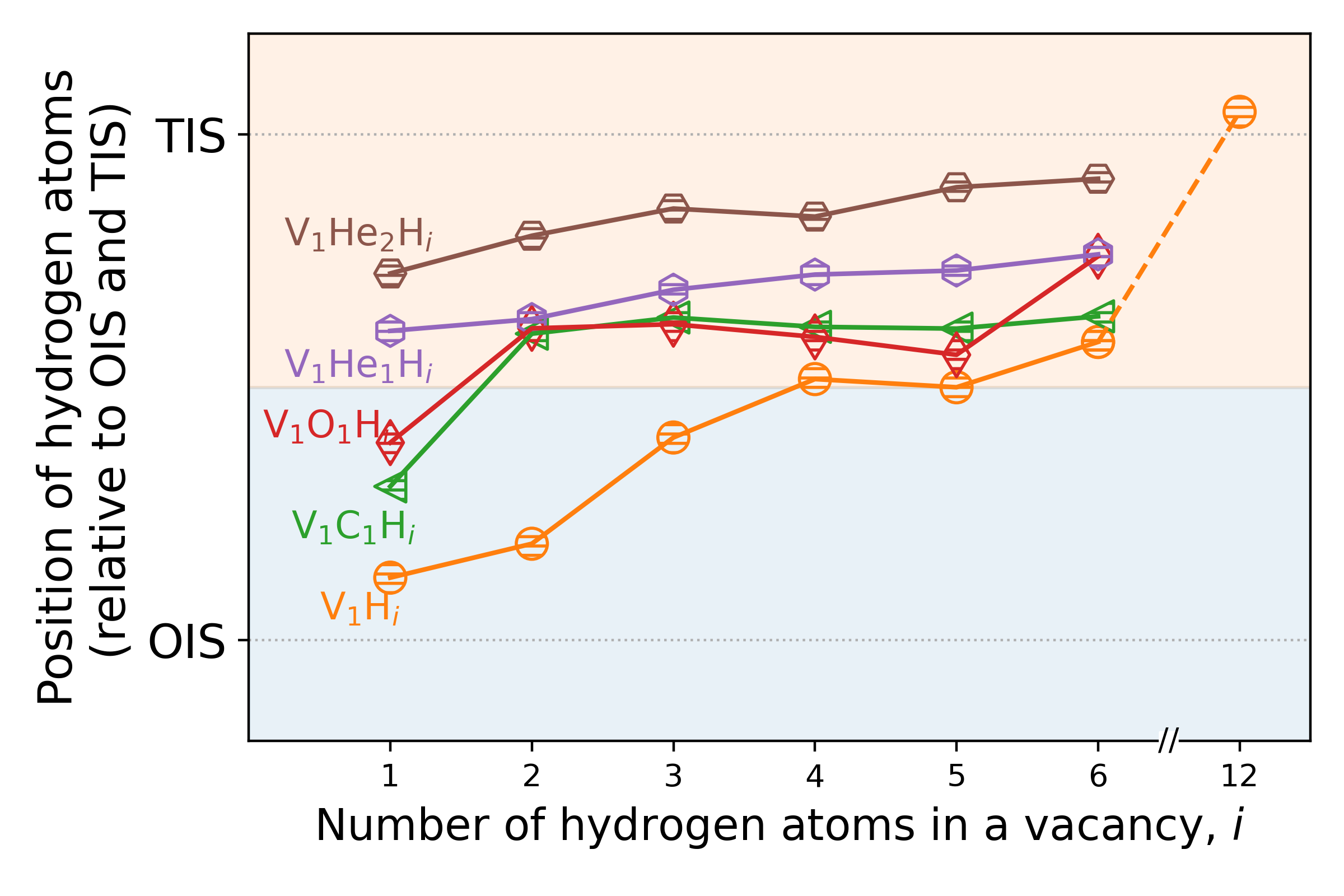}
  \caption{Average positions of hydrogen atoms (relative to the OIS and TIS) bonding to different types of vacancy complex.}
  \label{Fig:pos_evolve}
 \end{figure} 
 
  Fig. \ref{Fig:pos_evolve} shows the average positions of hydrogen atoms bonding to different types of vacancy complex. It can be clearly observed that the hydrogen atoms are closer to the TIS in vacancy complex containing impurities (including carbon, oxygen and helium) compared to those without impurities. Among these impurity atoms, the effect of helium atoms is most significant. The NRA/C results from the cases of V$_1$He$_1$D$_{4-6}$ (see Fig. \ref{Fig:ang_scan}n) exhibit best agreement with that of P-V experiments. The case of V$_1$He$_2$D$_1$ (see Fig. \ref{Fig:ang_scan}o) shows good agreement as well. However, the ratio of deuterium to vacancy is smaller than that predicted by SRIM and MD calculations presented in Sec.\ref{Sec:model_targ}. After taking into account the binding energy, the case of V$_1$He$_1$D$_6$ becomes less favorable, but the cases of V$_1$He$_1$D$_4$ and V$_1$He$_1$D$_5$ are still promising. Thus, we can predict that the signals of P-V experiments are mainly induced by vacancy complex containing both helium and hydrogen atoms. 
 
 \bigbreak
 
 Unlike carbon and oxygen which are intrinsic impurities in tungsten, the helium atoms should come from the probing beam of 750 keV $^3$He ions used in the NRA/C experiments. The authors of P-V experiments did not report the measurement fluence of $^3$He ions in their paper \cite{Picraux1974}. However, they reported a measurement fluence of $^3$He ions in another paper, which is $3 \times 10^{16}$ cm$^{-2}$ \cite{Picraux1974a}. In nowadays NRA experiment, the measurement fluence of $^3$He ions is on the order of magnitude of $1 \times 10^{15}$ cm$^{-2}$ \cite{Markelj2020}. Hence, we can estimate that the fluence of $^3$He ions is comparable to or even higher than that of deuterium ($3 \times 10^{15}$ cm$^{-2}$) in the P-V experiments. Due to high binding energies between helium and vacancies \cite{Becquart2007} and low migration energies of helium in tungsten ($\sim$ 0.06 eV \cite{Becquart2006, Zhou2010}), the helium from the measurement beam can be readily trapped by vacancies containing deuterium, resulting in the change of deuterium lattice locations. In fact, Picraux and Vook did observe that profiles of NRA/C angular scans can be affected by a continuous irradiation of 750 keV $^3$He ions in another experiment \cite{Picraux1975}. A discussion of this experiment is presented in next section.

\section{Discussion}

  In the previous section, we demonstrated that the contradictory results about the positions of deuterium/hydrogen in a vacancy indicated by the NRA/C experiments (close to the TIS) and that determined by DFT calculations (close to the OIS) can be resolved by considering the effect of helium atoms. In fact, in experimental conditions, there could be other potential factors affecting the positions of deuterium/hydrogen in a vacancy. It has been reported that the stable positions of hydrogen in a perfect tungsten can be influenced by anisotropic elastic strain \cite{Zhou2012}. Nonetheless, the effect of strain should be negligible due to the small amount of defects in the P-V experiments.
  
  \bigbreak  
  
  Another potential factor is the elevated temperature in the experiments. To this regard, we performed ab initio MD simulations at 300 K to investigate the movement of hydrogen atoms. The methods of ab initio MD simulations are given in supplementary materials. Fig. \ref{Fig:MD_H}a and Fig. \ref{Fig:MD_H}b show the probability density distributions of 1 and 12 hydrogen atoms bonding to a vacancy at 300 K, respectively. The simulation time is 4 ps. A warmer (colder) background color represents a higher (lower) possibility of finding hydrogen atoms at the corresponding position. For the cases of 1 and 12 hydrogen atoms, the hydrogen atoms are mainly vibrating around the OIS and TIS, respectively, which is in agreement with the DFT calculations. Hence, the elevated temperature at 300 K does not alter the stable positions of hydrogen in terms of the OIS and TIS. In addition, we can observe that, as shown in the right side of Fig. \ref{Fig:MD_H}b, some hydrogen atoms are leaving from the vacancy, implying that the filling level of hydrogen in a vacancy at room temperature is smaller than 12.   
  
  \begin{figure}[H]
  \centering
  \includegraphics[width=0.75\linewidth]{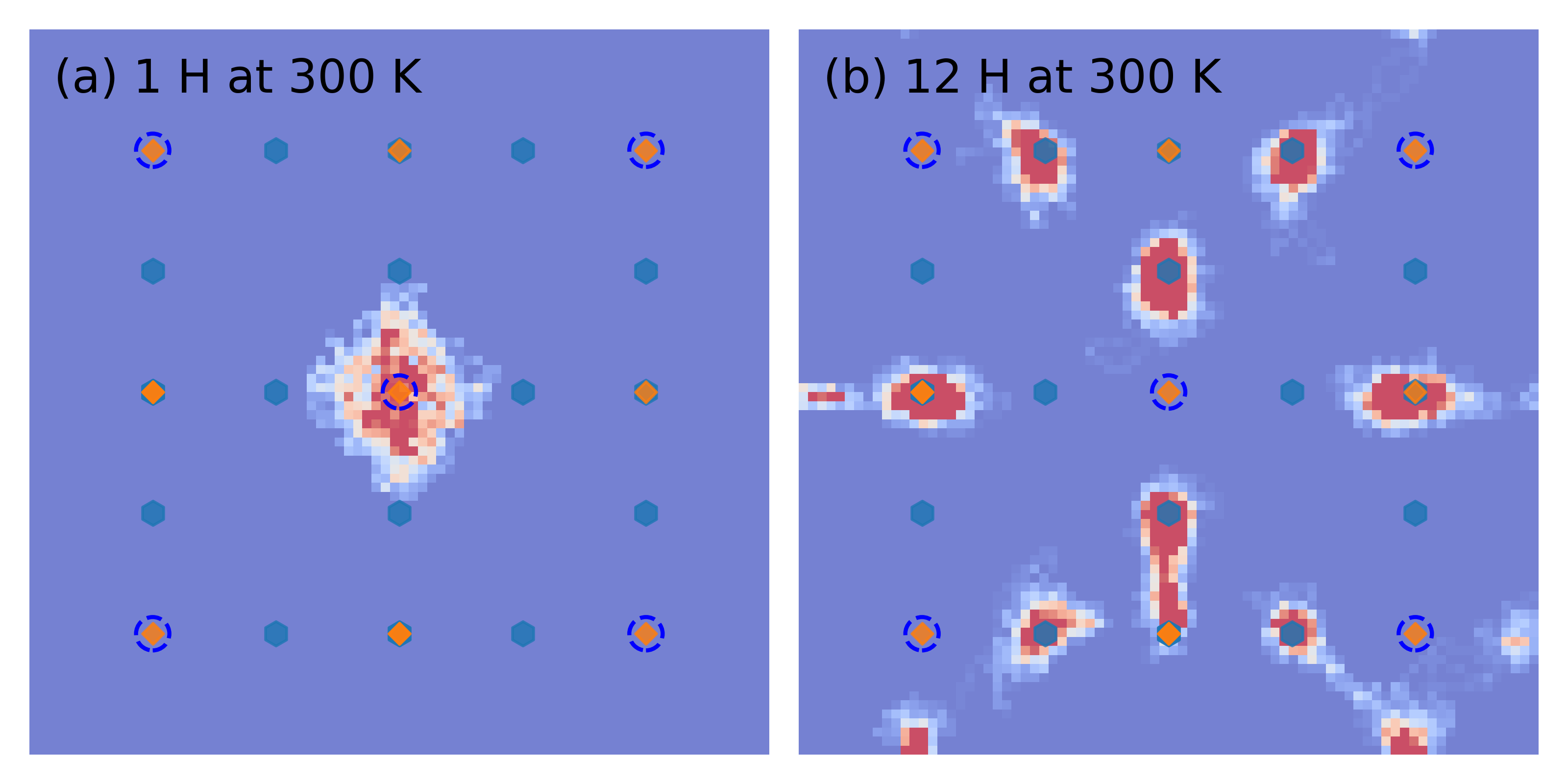}
  \caption{Probability density distributions of hydrogen atoms bonding to a vacancy at 300 K viewed along the [001] direction obtained from ab initio MD simulations for: (a) one hydrogen atom bonding to a vacancy and (b) 12 hydrogen atoms bonding to a vacancy. The probability of hydrogen at one position increases with a warmer background color. The locations of TIS (blue hexagon), OIS (orange square) and tungsten atoms in a perfect unit cell (blue circle) are indicated.}
  \label{Fig:MD_H}
 \end{figure}
 
 \bigbreak
 
 In addition to the impact of helium, another major result from the study of deuterium lattice locations is related to the deuterium filling level. To collaborate and confront with NRA/C, we performed macroscopic rate equation (MRE) modelling to investigate the population of V$_1$D$_i$ under the P-V experimental conditions prior to the application of $^3$He ion beam. The initial depth distribution of deuterium and the depth profile of vacancies follow the ones obtained from the SRIM and MD calculations shown in Fig. \ref{Fig:srim_profile}. The deuterium diffusion coefficient is taken from previous DFT calculations \cite{Fernandez2015} with a diffusion barrier energy of 0.2 eV. The detrapping energies of deuterium from the vacancies were set to the binding energies given in Fig. \ref{Fig:bind_ener} plus the diffusion barrier energy. 30 keV deuterium atoms were implanted to tungsten with a flux of $3 \times 10^{13}$ cm$^{-2}$s$^{-1}$. After the fluence reached the P-V experimental one, the implantation was finished and the modelling entered to a resting period for 100 s at 296 K. Detailed mechanisms of the MRE modelling implemented in the MHIMS-R code can be found in previous works \cite{Hodille2016}.
 
 \begin{figure}[H]
  \centering
  \includegraphics[width=0.65\linewidth]{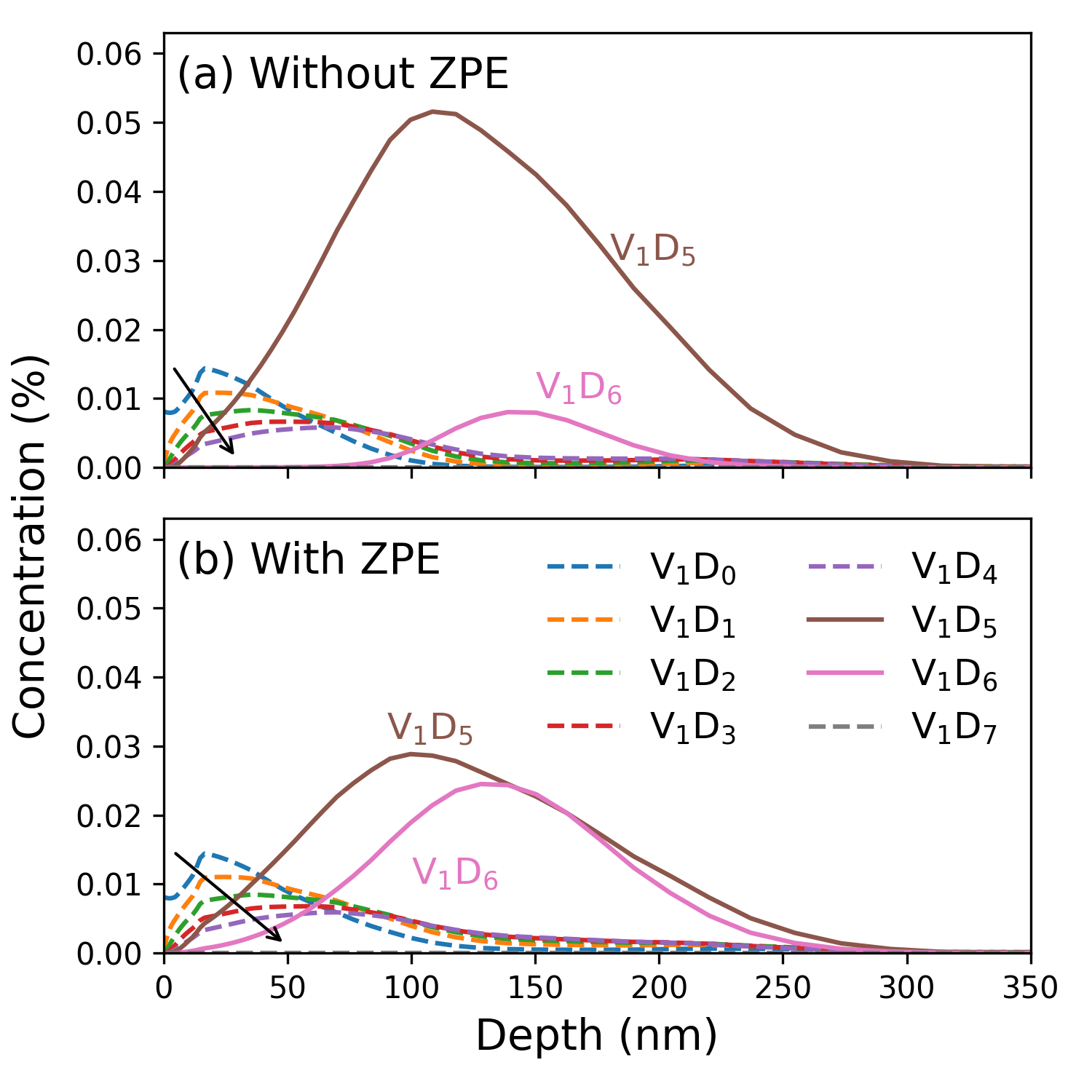}
  \caption{Concentrations of V$_1$D$_i$ under the P-V experimental conditions obtained from the rate equation modelling using binding energies (a) without the ZPE correction and (b) with the ZPE correction. (The arrows indicate an increase of filling level.)}
  \label{Fig:vac_concen}
 \end{figure}
 
 \bigbreak
 
 Fig. \ref{Fig:vac_concen}a shows the concentrations of V$_1$H$_i$ obtained from the MRE modelling after the resting period, in which the binding energies without the ZPE correction are used. We can observe that the dominant vacancy complex is V$_1$H$_5$. After taking into account the ZPE correction, as shown in Fig. \ref{Fig:vac_concen}b, the concentrations of V$_1$H$_5$ and V$_1$H$_6$ are on a similar level. The vacancy complexes with a filling level smaller than 5 are mainly populated at the near surface region. Whereas, there is almost no vacancy complex with more than 6 deuterium present in the target. In terms of the scenarios of V$_1$He$_1$H$_i$, our preliminary MRE modelling shows that the population of V$_1$He$_1$H$_6$ significantly diminishes due to lower binding energy so that the only dominant structure is V$_1$He$_1$H$_5$. Thus, we can find that the deuterium filling levels obtained from the NRA/C and MRE are in good agreement.  
 
 \bigbreak
 
\begin{figure}[H]
  \centering
  \includegraphics[width=0.75\linewidth]{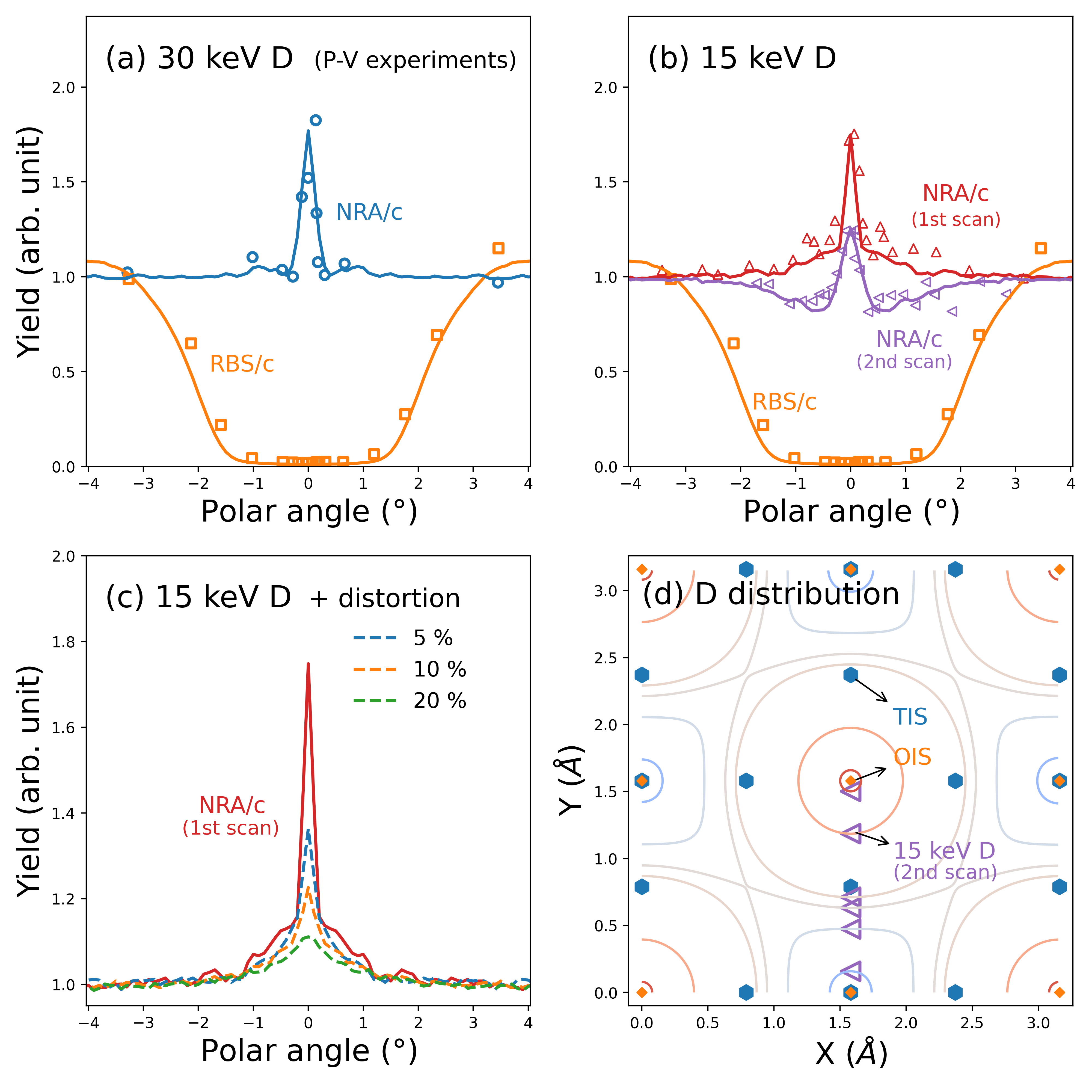}
  \caption{Analysis of positions of deuterium in different irradiation conditions: Experimental (dots) and simulated (lines) NRA/C and RBS/c angular scans for (a) 30 keV deuterium implanted to tungsten (the P-V experiments) and (b) 15 keV deuterium implanted to tungsten under low and high radiation damage induced by 750 keV $^3$He ions, (c) simulated NRA/C angular scans for 15 keV deuterium implanted to tungsten with lattice distortion ranging from 5 \% to 20 \% (dashed lines) and (d) illustration of deuterium positions that are used to fit NRA/C results with high damage shown in (b). The angular scans are through the [001] axis. (Experimental values in (a) and (b) are extracted from ref \cite{Picraux1974} and ref \cite{Picraux1975}, respectively.)}
  \label{Fig:ang_scan_He}
 \end{figure}

 We have shown how different trapping conditions of deuterium atoms affects the deuterium lattice locations in the previous section. In turn, a variation of NRA/C results can indicate a change of deuterium lattice locations due to the evolution of defects. Fig. \ref{Fig:ang_scan_He}a shows the comparison of simulated NRA/C and RBS/c results to those of P-V experiments, in which the simulated NRA/C signals are generated from V$_1$He$_1$H$_5$. Fig. \ref{Fig:ang_scan_He}b shows another NRA/C experiments performed by Picraux and Vook \cite{Picraux1975}, in which 15 keV deuterium with a fluence of $1 \times 10^{15}$ cm$^{-2}$ were implanted to tungsten at room temperature. The probing ions are still 750 keV $^3$He ions. Initial results of NRA/C angular scan (red triangles) exhibit a central peak and prominent shoulder signals. The best agreement with the experiments is obtained from simulations (red line) generated from V$_1$He$_3$D$_3$. This indicates that the deuterium lattice locations are already disturbed by the helium atoms. After the initial angular scan, Picraux and Vook performed another angular scan following additional irradiation of 750 keV $^3$He ions with a fleunce around $3.7 \times 10^{16}$ cm$^{-2}$ \cite{Picraux1975}. The experimental results of this second angular scan (purple triangles in Fig. \ref{Fig:ang_scan_He}b) exhibit significant difference with the first one. The yield of central peak decreases and signals of shoulder peaks are replaced by dips.
 
 \bigbreak
 
 To study the trapping condition in the second angular scan shown in Fig. \ref{Fig:ang_scan_He}b, we simulated NRA/C signals by increasing the number of helium atoms in a vacancy up to 6. However, the simulated results cannot get a satisfactory agreement with the experimental ones. Furthermore, the binding energies of hydrogen atoms decrease with an increase of helium atoms (see Fig. \ref{Fig:bind_ener}), implying an easier detrapping of hydrogen atoms at room temperature. If fact, it has been postulated the evolution of NRA/C angular scans with additional helium irradiation can either be attributed to distortion on probing ions induced by the presence of large amount of helium atoms in tungsten \cite{Picraux1974a} or due to an evolution of defects trapping deuterium atoms \cite{Picraux1981}.
 
 \bigbreak
 
 We studied the effect of distortion of probing ions on the first angular scan in Fig. \ref{Fig:ang_scan_He}b. The lattice locations of deuterium were kept fixed, but 5 \% to 20 \% of tungsten atoms were randomly displaced which can distort spatial distribution of probing ions in channeling mode. As shown in Fig. \ref{Fig:ang_scan_He}c, with a higher degree of distortion, the NRA/C central peak (dashed lines) diminishes, but there is no sign of formation of dips as in the case of the second scan in Fig. \ref{Fig:ang_scan_He}b. Hence, an evolution of trapping defects should be responsible for the change of NRA/C signals in Fig. \ref{Fig:ang_scan_He}b. 
 
  \bigbreak 
 
  We found that the results of second angular scan can be fitted by simulations in which deuterium atoms (purple markers) are placed at locations as indicated in Fig. \ref{Fig:ang_scan_He}d. Since the spatial distribution of probing ions in a channeling mode is determined by a so-called continuum potential \cite{Lindhard1965}, deuterium atoms can be put at other positions with the same continuum potential which are indicated by the lines crossing deuterium atom shown in Fig. \ref{Fig:ang_scan_He}d. A general feature of the deuterium location is that the deuterium atoms are spread between the OIS and TIS. This feature may imply an evolution of defect types from isolated to extended ones, such as vacancy clusters providing large space for the decoration of deuterium atoms. One plausible mechanism for the formation of vacancy clusters can be trap mutation induced by large amounts of helium atoms \cite{Boisse2014}. DFT studies have shown that the stable positions of hydrogen atoms in di-vacancies are still close to the OIS \cite{Yang2018}. Thus, we can expect that there could be larger extended defects. The cases of deuterium trapped by extended defects will be studied in the future.
  
  \bigbreak
  
  The above studies suggest that it is important to have a clear record of the probing helium ions when one performs NRA/C experiments. This is not only to make the experiments possible to reproduce or simulate, but also to take into account the effect of helium atoms on defect evolution. For the design of NRA/C experiments, it would be preferable to increase the ratio between the total amount of deuterium and helium atoms. 

\section{Conclusions}

In this work, we studied the lattice locations of deuterium in tungsten and the correlation of the lattice locations with trapping conditions. DFT calculations were used to determine the lattice locations of hydrogen at different trapping conditions. A newly developed NRA/C simulation software was applied to generate signals of NRA/C angular scans through the [001] axis of tungsten. We systematically produced the NRA/C signals for deuterium retained in perfect tungsten and trapped by defects of interstitial and vacancy type at different deuterium filling levels. 

\bigbreak

Our studies show that the trapping conditions affect the relative distance of deuterium to the OIS and TIS sites, resulting in distinct features in NRA/C signals.
The comparison between NRA/C simulations and available experiments \cite{Picraux1974,Picraux1975} suggests that the deuterium atoms were trapped by vacancies, and their lattice locations is affected by helium from the probing beam used for NRA/C. Comparing with the DFT calculations, we find that the P-V experiments can be reproduced by the majority of deuterium having been trapped by the vacancy complexes V$_1$He$_1$H$_4$ and/or V$_1$He$_1$H$_5$. This prediction is in a good agreement with that obtained from macroscopic rate equation modelling, which favors the V$_1$He$_1$H$_5$ vacancy complex.

\bigbreak

More generally, our study demonstrates that by combining the NRA/C and DFT methods one can deduce the atomic nature of crystal structures trapping deuterium. This opens up the possibility to use further NRA/C experiments for different materials and/or hydrogen loading conditions to gain additional insight on hydrogen behaviour in metals.

  \section*{Acknowledgements}

  X. Jin acknowledges Thomas Schwarz-Selinger, Huan Liu, Ilja Makkonen and Tommy Ahlgren for fruitful discussions. This work has been carried out within the framework of the EUROfusion Consortium, funded by the European Union via the Euratom Research and Training Programme (Grant Agreement No 101052200 — EUROfusion). Views and opinions expressed are however those of the author(s) only and do not necessarily reflect those of the European Union or the European Commission. Neither the European Union nor the European Commission can be held responsible for them. The authors wish to acknowledge CSC – IT Center for Science, Finland, for computational resources.

  \section*{Competing interests}

  The authors declare no conflict of interest.

\bibliographystyle{naturemag}
\bibliography{Manuscript.bib}

\end{document}


\maketitle

\section{Comparisons of NRA simulations and experiments}

  We performed comparisons of NRA simulations to NRA experiments utilizing 1500 keV and 2000 keV deuterium to probe pure carbon using a reaction of $^{12}$C(d,p)$^{13}$C \cite{Simpson1986} (Note that the NRA simulations and experiments in this case are not in the channeling mode). In our simulations, all carbon atoms in carbon targets were randomly displaced. The thickness of carbon targets was set to 15 and 20 µm for simulations using 1500 keV and 2000 keV deuterium, respectively. The Debye temperature of carbon at room temperature was set to 1507 K \cite{Tohei2006}, resulting a one dimensional root mean square thermal vibration magnitude of 4 pm. Proton generated from the nuclear reaction was detected by a detector located at 135$^{\circ}$. Differential cross sections of $^{12}$C(d,p)$^{13}$C measured at 135$^{\circ}$ \cite{Csedreki2014} were used. The multiple scattering effect was taken into account by setting the spread angle of emitting protons to 1$^{\circ}$. 
  
  \bigbreak
  
\begin{figure}[H]
  \centering
  \includegraphics[width=0.75\linewidth]{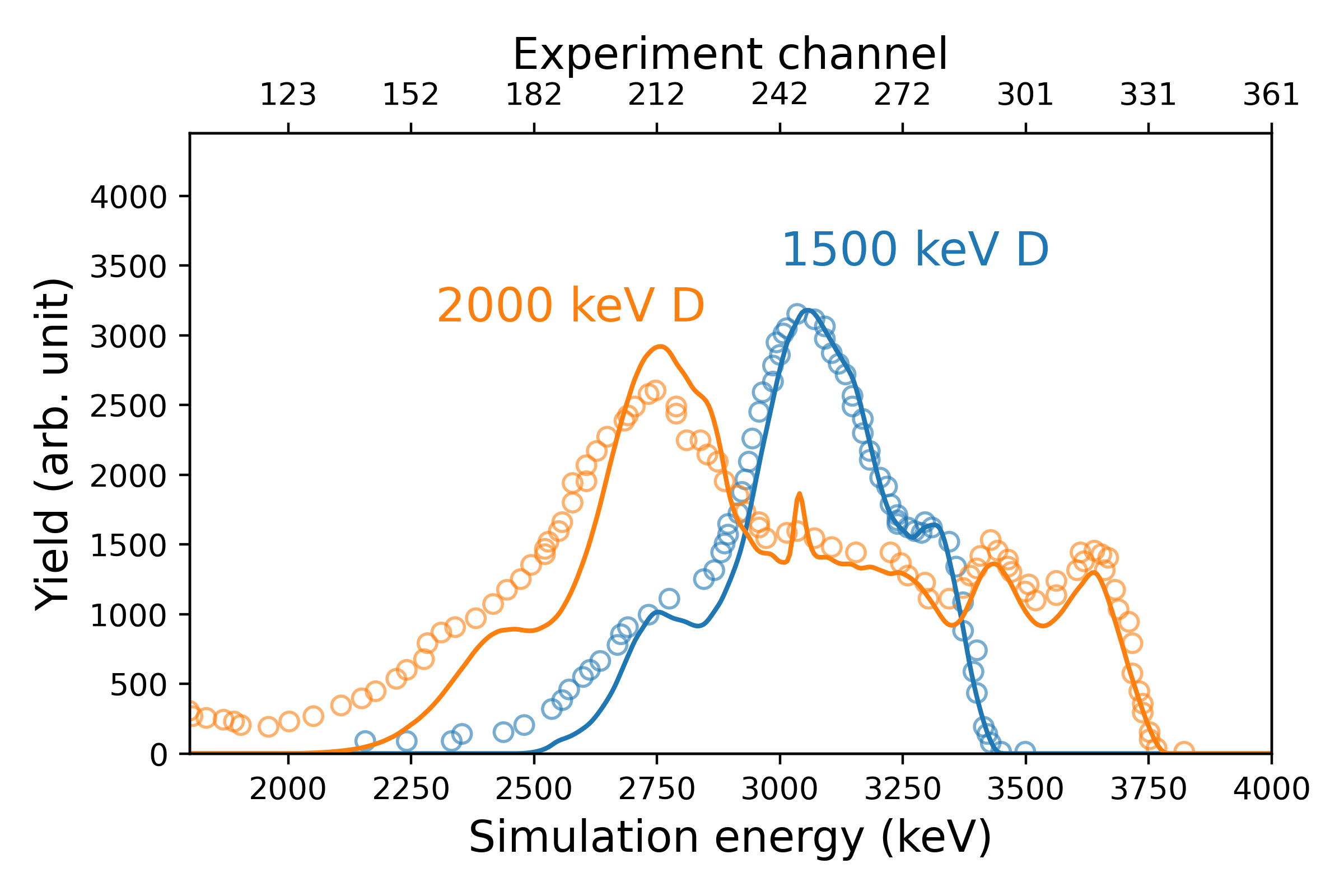}
  \caption{NRA spectra of 1500 keV (blue) and 2000 keV (orange) deuterium on carbon: simulation and experimental spectra are represented by lines and circles, respectively. (Experimental data are extracted from ref.\cite{Simpson1986}.)}
  \label{Fig:H_pos_int}
\end{figure}

  Fig.\ref{Fig:H_pos_int} shows the comparisons of the simulated (lines) and experimental (circles) NRA spectra using 1500 keV (blue) and 2000 keV (orange) deuterium. The agreement between the simulation and experiments is satisfactory especially at high energy regions.
  
\section{Comparison of range distributions in amorphous, polycrystalline and crystalline tungsten}

In single crystalline materials and under some irradiation conditions (combination of ion species, energy, incoming angle, and target material),  ion channeling effects \cite{Oen63c,Gem74,Hob95} may dramatically enhance ion ranges compared to those in an amorphous materials. Regarding pure elemental metals, this issue is particularly important since these cannot be amorphized \cite{Bul56}, and typical conventional metals are polycrystalline with randomly oriented grains of sizes in the tens or hundreds of micrometers size range \cite{Callister}. Since the ranges of keV ions are in the nanometer regime, in effect each ion impacts on a single crystalline grain of random surface orientation. We have recently shown that under some irradiation conditions, a large fraction of all ions incoming also on a polycrystalline material may undergo channeling, leading to range distributions that strongly differ from that in amorphous materials \cite{Nor16}. Hence we here consider whether channeling may affect also the range profile under the current irradiation condition of 30 keV D ions incoming on tungsten.

\bigbreak

\begin{figure}[H]
  \centering
  \includegraphics[width=0.95\linewidth]{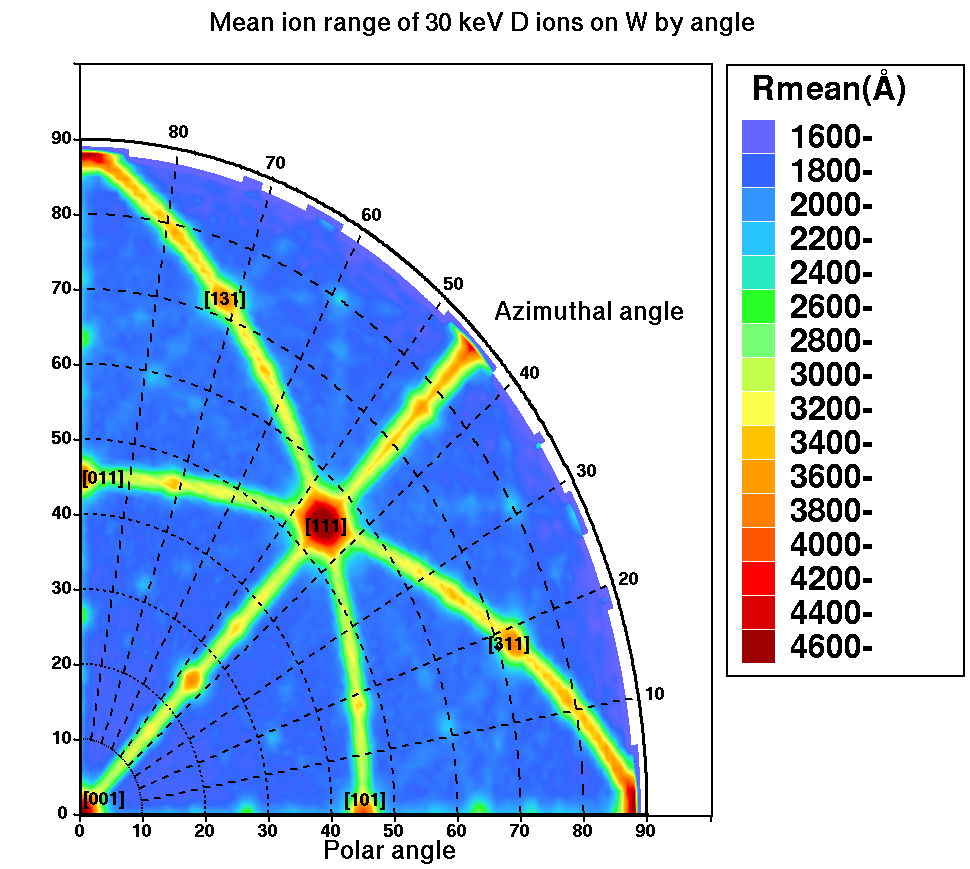}
  \caption{Channeling map \cite{Nor16} for 30 keV D ions in W. The colors indicate the mean projected range of ions in each direction according to the scale on the right. 
  }
  \label{Fig:channeling_map}
\end{figure}

Since the SRIM code can only handle amorphous materials, we used the MDRANGE code \cite{Nor94b,Nor16,MDRANGE} to examine channeling into crystalline materials. The simulations were set up following the principles described in Ref. \cite{Nor16}. To get a systematic view of in which directions channeling is important, we calculated the channeling map of mean ion ranges. The map is presented in Fig.\ref{Fig:channeling_map} and shows that there is strong channeling in the principal directions 100, 110 and 111, as well as weaker axial and planar channeling in several other directions. The angular channel widths are $\lesssim 5^\circ$, and there are large areas in the map with practically no channeling.

\bigbreak

To understand how the channeling affects range distributions, we also simulated range profiles of ions implanted into a few selected directions, namely the principal crystal axes 100, 110 and 111, as well testing how tilting the beam off the 100 axis affects the results. We also mimicked irradiation of a ''polycrystalline'' sample by simulating projected ranges selected randomly over all crystallographically inequivalent directions in a cubic crystal system, i.e. within the triangle spanned by the 100 - 110 -111 directions. To enable comparison
with SRIM \cite{SRIM-2013}, we also simulated ranges into a hypothetical amorphous W sample (generated by setting atom coordinates in random positions in a (4 nm)$^3$ cell with a minimum distance of 2.1 \AA{} between the atoms). Both the SRIM and MDRANGE simulations used the universal Ziegler-Biersack-Littmark repulsive potential \cite{ZBL} and SRIM-2013 electronic stopping power
\cite{SRIM-2013}.

\begin{table}[]
    \centering
    \begin{tabular}{|l|l|l|c|c|c|c|}
        \hline
         Method & Sample & Surface & $\theta$ & $\phi$ & Mean range & Straggle \\
         & structure & orientation & $(^\circ)$ & $(^\circ)$ & (Å) & (Å) \\
        \hline
         SRIM    & Amorphous & --  & 0 & 0 & 1600 $\pm$ 01 & 708 \\
         MD & Amorphous & -- & 0 & 0 & 1550 $\pm$ 5 & 696 \\
         MD & Polycrystalline & random & 0* & 0* & 2040 $\pm$ 7 & 1030 \\
         MD & Single crystalline & 100 & 7 & 0 & 1922 $\pm$ 6 & 942 \\
         MD & Single crystalline & 100 & 7 & random & 2010 $\pm$ 7 & 1070\\      
         MD & Single crystalline & 100 & 7 & 27 & 1770 $\pm$ 6 & 933 \\
         MD & Single crystalline & 100 & 20 & 20 & 1651 $\pm$ 6 & 821 \\
         MD & Single crystalline & 100 & 0 & 0 & 4670  $\pm$ 10 & 900\\
         MD & Single crystalline & 110 & 0 & 0 & 4080  $\pm$ 10 & 1020\\
         MD & Single crystalline & 111 & 0 & 0 & 4780  $\pm$ 10 & 730\\
         \hline

    \end{tabular}
    \caption{Mean ion ranges and stragglings (standard deviation of the distribution) for some individual implantation conditions. The MD calculations were carried out with the MDRANGE code that simulates ion ranges in the recoil interaction approximation of MD. The error bar is the standard error of the mean. The error for the straggle is approximately $1/2$ of the error of the mean range. The asterisk (*) for the polycrystalline sample indicates that simulation of the polycrystalline material was in practise implemented by scanning over different angles and using the projected range (see text).
    }
    \label{tab:meanranges}
\end{table}

  \bigbreak

\begin{figure}[H]
  \centering
  \includegraphics[width=0.8\linewidth]{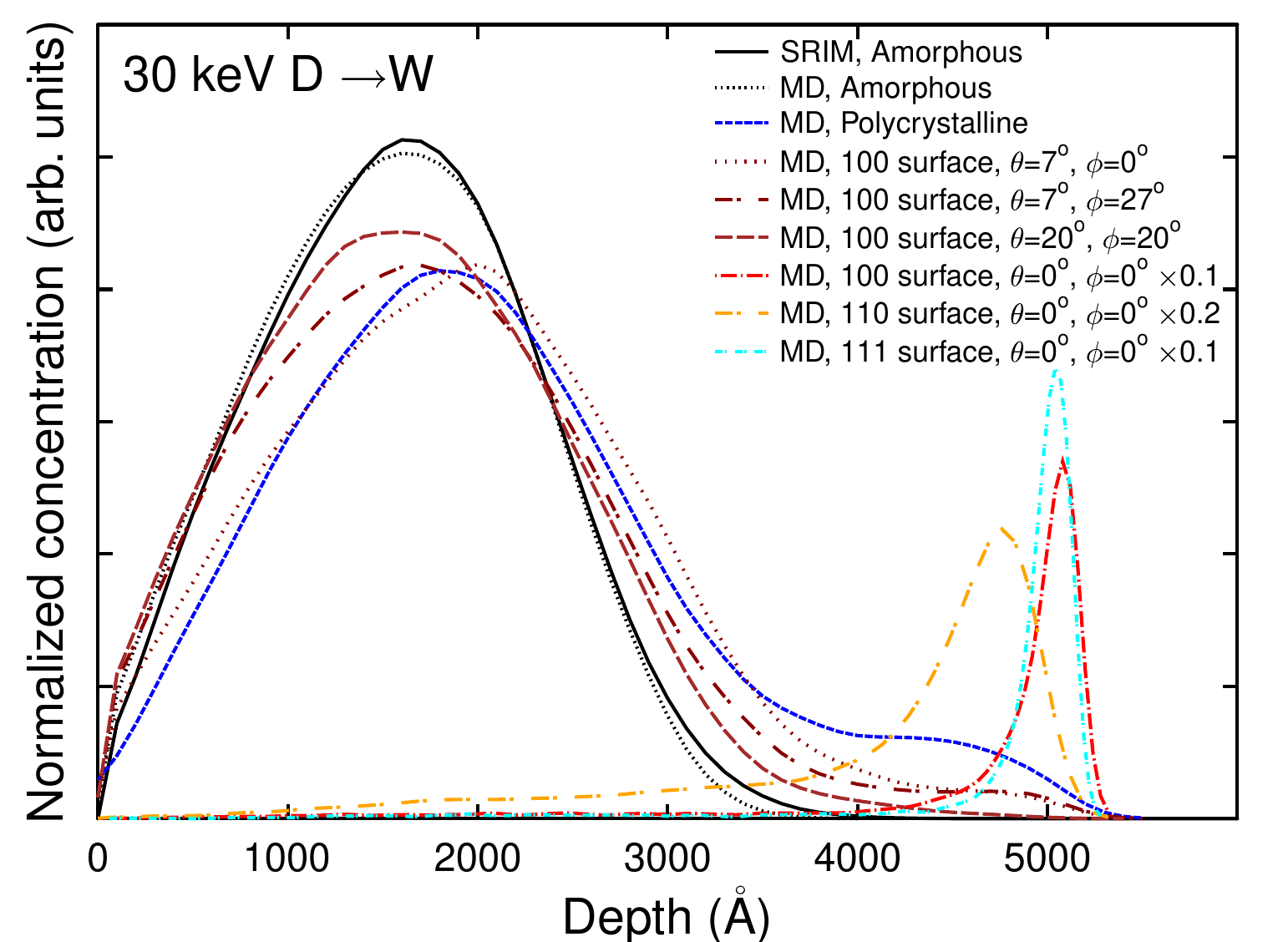}
  \caption{Range distributions of 30 keV D ions in W calculated with SRIM and the MD code MDRANGE, for various sample structures and incoming angles. The distributions are smoothed with a Gaussian function and normalized to the same area. Since the range distributions into the strong channeling directions 100, 110 and 111 are strongly peaked, these are multiplied by a factor $< 1$ for clarity in the plotting.}
  \label{Fig:ranges}
\end{figure}

The results in Fig. \ref{Fig:ranges} and Table \ref{tab:meanranges} show that, as expected, the SRIM and MDRANGE results agree very well for amorphous materials. All varieties of the simulations of crystalline structures are somewhat different. As expected from the channeling map, implantation directly into the low-index crystal directions 100, 110 and 111 have much larger ranges than the amorphous material. Moreover, since there is a fairly wide range of channeling directions (Fig. \ref{Fig:channeling_map}), it is not surprising that the results for the model 'polycrystalline' implantation condition also shows a channeling tail. However, also the case of tilting the beam 7 degrees in the polar angle off the 100 direction, without any tilting of the azimuthal angle, shows a clear channeling tail. This can be understood in terms of the planar channeling streak seen in Fig. \ref{Fig:channeling_map}.

\bigbreak

The cases where the beam is tilted both in polar and azimuthal angle ($\theta = 7^\circ, \phi=27^\circ$ and $\theta = 20^\circ, \phi=20^\circ$)
show only a weak tail resulting from secondary channeling (ions that happen to scatter into a channeling direction). In these cases, the mean ranges differ only some $\sim$ 10\% from that in the hypothetical amorphous material.

\section{Deuterium/Hydrogen bonding to vacancies with 3 to 6 helium atoms}

\begin{figure}[H]
  \centering
  \includegraphics[width=1.0\linewidth]{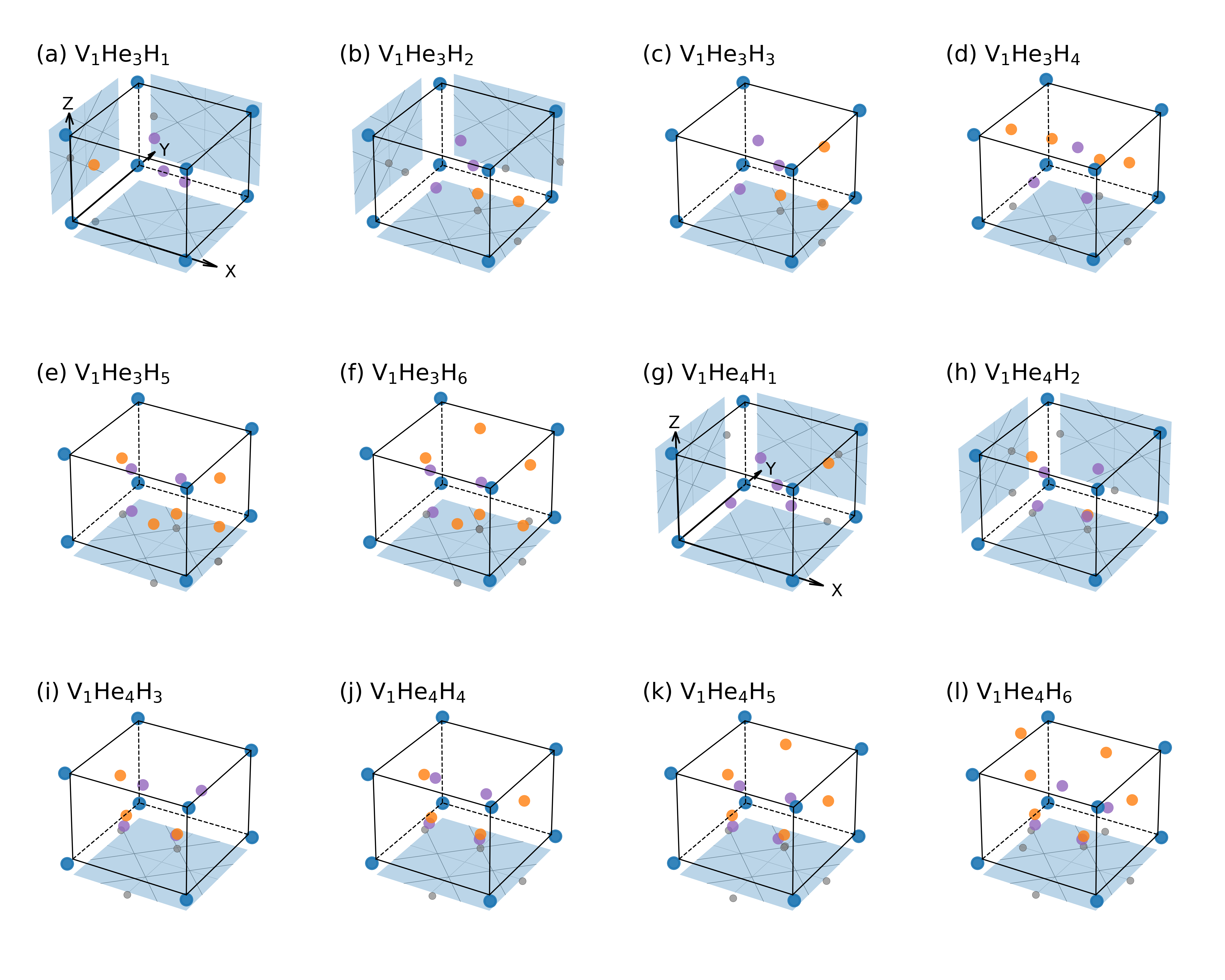}
  \caption{Positions of hydrogen atoms (orange) in a vacancy containing 3 helium atoms (a-f) and containing 4 helium atoms (g-l). The helium atoms are represented by the purple circles. Other notations are same to those in the manuscript. (The projections to the (100) and (010) planes are not shown for vacancies with more than 2 hydrogen atoms for the sake of simplicity.)}
  \label{Fig:pos_v1He3He4}
\end{figure}

\begin{figure}[H]
  \centering
  \includegraphics[width=1.0\linewidth]{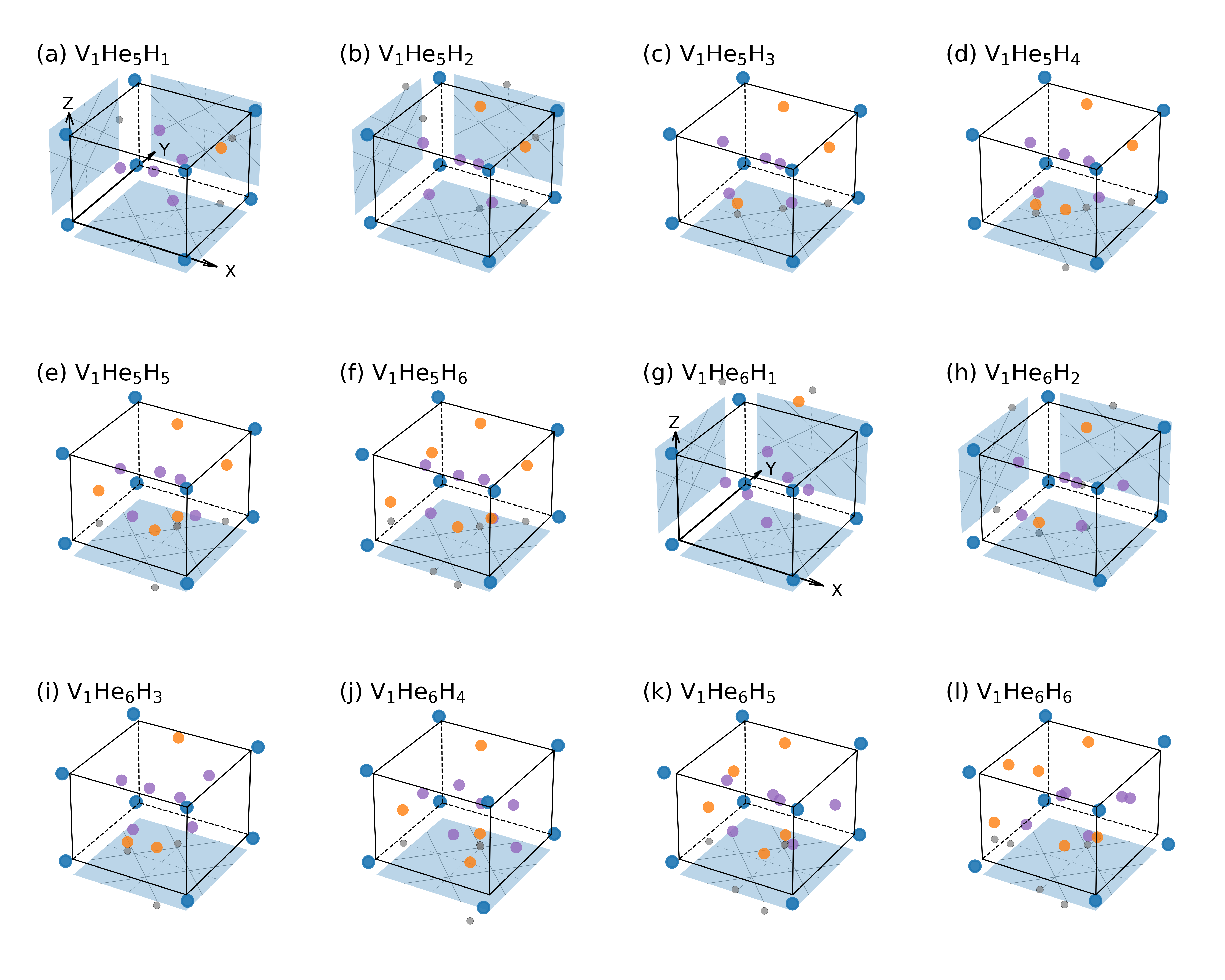}
  \caption{Positions of hydrogen atoms (orange) in a vacancy containing 5 helium atoms (a-f) and containing 6 helium atoms (g-l). The helium atoms are represented by the purple circles. Other notations are same to those in the manuscript. (The projections to the (100) and (010) planes are not shown for vacancies with more than 2 hydrogen atoms for the sake of simplicity.)}
  \label{Fig:pos_v1He5He6}
\end{figure}

\begin{figure}[H]
  \centering
  \includegraphics[width=1.0\linewidth]{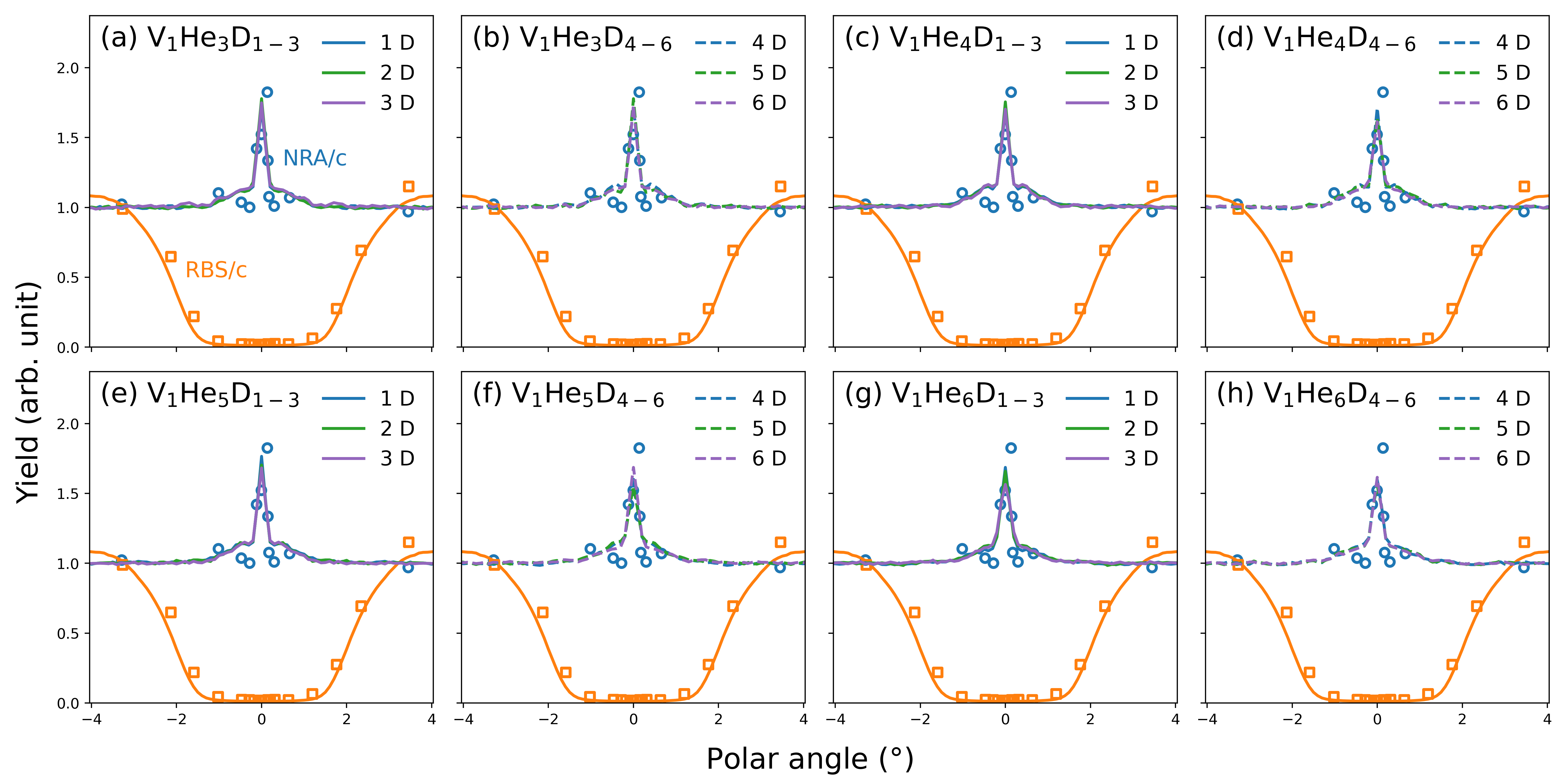}
  \caption{Comparison of experimental (dots) and simulated (lines) RBS/c and NRA/c angular scans through the [001] axis of tungsten with deuterium at different conditions: (a-b) bonding to a vacancy containing 3 helium atoms, (c-d) bonding to a vacancy containing 4 helium atoms, (e-f) bonding to a vacancy containing 5 helium atoms, (g-h) bonding to a vacancy containing 6 helium atoms. (Experimental values are extracted from the P-C experiments \cite{Picraux1974}.)}
  \label{Fig:ang_scan_bis}
\end{figure}

  The locations of hydrogen atoms in V$_1$He$_{3-6}$H$_i$ are shown in Fig.\ref{Fig:pos_v1He3He4}(a-f), Fig.\ref{Fig:pos_v1He3He4}(g-l), Fig.\ref{Fig:pos_v1He5He6}(a-f) and Fig.\ref{Fig:pos_v1He5He6}(g-l), respectively. In general, the hydrogen atoms are close to the TIS. With a higher number of helium atoms, some hydrogen atoms are positioned outside the vacancy surface. The simulated NRA/c results are presented in Fig.\ref{Fig:ang_scan_bis}, which are featured with two shoulder peaks on the two sides of a central peak. The NRA/c results of V$_1$He$_3$H$_{1-3}$, as shown in Fig.\ref{Fig:ang_scan_bis}a, is almost same with that of deuterium in perfect tungsten. When the number of helium atoms is increased, we can observe that the yield of central peak gradually decreases, which can be attributed to a larger distance between deuterium atoms and the surface of vacancy.  

\section{Methods of ab initio molecular dynamics simulations}

  Ab initio MD simulations were performed with the Vienna ab initio simulation package (VASP) \cite{Kresse1993, Kresse1994, Kresse1996} using projector augmented wave potentials \cite{Blochl1994, Kresse1999}. The electron exchange correlation was described with generalized gradient approximation (GGA) using Perdew-Burke-Ernzerhof (PBE) functionals \cite{Perdew1996}. As in the DFT calculations, we employed a $3 \times 3 \times 3$ supercell composed of 54 lattice points, and applied a $5 \times 5 \times 5$ $k$-point gird obtained using the Monkhorst-Pack method \cite{Monkhorst1976}. A plane wave energy cutoff of 350 eV was used. We used a canonical ensemble with a Nose-Hoover thermostat. We performed the simulations at 300 K for 4 ps with a time step of 1 fs. Initial configurations of hydrogen atoms in vacancies are at stable positions. 

\bibliographystyle{naturemag}
\bibliography{Supplementary_materials.bib}
